\documentclass[10pt,twoside]{IEEEtran}

\usepackage{amsfonts,graphicx,subfigure}
\usepackage{amsfonts,amsmath,amssymb}
\usepackage{epic,eepic,eepicemu}
\usepackage{epsf}
\usepackage{fancyheadings}
\usepackage{graphics}
\usepackage{psfrag}

\makeatletter
\long\def\@makecaption#1#2{
        \vskip 0.8ex
        \setbox\@tempboxa\hbox{\small {\bf #1:} #2}
    \parindent 1.5em  
        \dimen0=\hsize
        \advance\dimen0 by -3em
        \ifdim \wd\@tempboxa >\dimen0
                \hbox to \hsize{
                        \parindent 0em
                        \hfil
                        \parbox{\dimen0}{\def\baselinestretch{0.96}\small
                                {\bf #1.} #2
                                }                         \hfil}
        \else \hbox to \hsize{\hfil \box\@tempboxa \hfil}
        \fi
        }
\makeatother

\renewcommand{\baselinestretch}{1}

\newcommand{\clset}{\ensuremath{C}}

\newcommand{\aseta}{\ensuremath{A}}

\newcommand{\myproj}[2]{\ensuremath{{#1}({#2})}}

\newcommand{\treedist}{\ensuremath{\rho}}
\newcommand{\treedistvec}{\ensuremath{{\boldsymbol{\rho}}}}
\newcommand{\trdistvec}{\ensuremath{{\boldsymbol{\rho_e}}}}

\newcommand{\grmaxmarg}{\ensuremath{\nu}}
\newcommand{\grmaxmargb}{\ensuremath{\nu}}

\newcommand{\spnodenum}{\ensuremath{n}}

\newcommand{\stsp}{\ensuremath{\statesp^{\spnodenum}}}

\newcommand{\Locset}{\ensuremath{\operatorname{LOCAL}}}

\newcommand{\treegr}{\ensuremath{T}}

\newcommand{\tausym}{\ensuremath{\tau}}
\newcommand{\tausymb}{\ensuremath{\tau}}

\newcommand{\meanpar}{\ensuremath{\mu}}
\newcommand{\meanparb}{\ensuremath{\mu}}

\newcommand{\taupar}{\ensuremath{\tau}}

\makeatletter
\long\def\@makecaption#1#2{
        \vskip 0.8ex
        \setbox\@tempboxa\hbox{\small {\bf #1:} #2}
        \parindent 1.5em  
        \dimen0=\hsize
        \advance\dimen0 by -3em
        \ifdim \wd\@tempboxa >\dimen0
                \hbox to \hsize{
                        \parindent 0em
                        \hfil 
                        \parbox{\dimen0}{\def\baselinestretch{0.96}\small
                                {\bf #1.} #2
                                } 
                        \hfil}
        \else \hbox to \hsize{\hfil \box\@tempboxa \hfil}
        \fi
        }
\makeatother

\long\def\comment#1{}

\makeatletter
\def\@cite#1#2{[\if@tempswa #2 \fi #1]}
\makeatother

\makeatletter
\long\def\barenote#1{
    \insert\footins{\footnotesize
    \interlinepenalty\interfootnotelinepenalty 
    \splittopskip\footnotesep
    \splitmaxdepth \dp\strutbox \floatingpenalty \@MM
    \hsize\columnwidth \@parboxrestore
    {\rule{\z@}{\footnotesep}\ignorespaces
      #1\strut}}}
\makeatother

\newcommand{\Ind}{\ensuremath{\mathbb{I\,}}}

\newcommand{\Margset}{\ensuremath{\operatorname{MARG}}}

\newcommand{\xoptset}{\ensuremath{\operatorname{OPT}}}

\newcommand{\eps}{\ensuremath{\epsilon}}

\newcommand{\bit}{\begin{itemize}}
\newcommand{\eit}{\end{itemize}}
\newcommand{\ben}{\begin{enumerate}}
\newcommand{\een}{\end{enumerate}}

\newcommand{\bear}{\begin{eqnarray}}
\newcommand{\eear}{\end{eqnarray}}

\newcommand{\vtiny}{\vspace*{.1in}}

\newcommand{\fn}{\footnotesize}

\newcommand{\supp}{\ensuremath{\operatorname{supp}}}

\newcommand{\clipotvec}{\ensuremath{{\boldsymbol{\clipot}}}}

\newcommand{\treepoly}{\ensuremath{\mathbb{T}}}

\newcommand{\tract}{\mathfrak{T}}

\newcommand{\cpar}{\ensuremath{\beta}}

\newcommand{\lagmul}{\ensuremath{\lambda}}

\newcommand{\Lag}{\ensuremath{{\mathcal{L}}}}

\newcommand{\Partinf}{\ensuremath{\Part_{\infty}}}

\newcommand{\df}{\ensuremath{d}}

\newcommand{\alphnorm}{\ensuremath{\kappa}}

\newcommand{\eparams}{{\ensuremath{\eparam^*}}}

\newcommand{\eparambar}{\ensuremath{\bar{\eparam}}}
\newcommand{\eparamtil}{\ensuremath{{\wtil{\eparam}}}}

\newcommand{\sumind}{\ensuremath{\alpha}}

\DeclareMathOperator{\suchthat}{s.t.}

\newcommand{\statenum}{{\ensuremath{m}}}

\newcommand{\tind}{\ensuremath{i}}

\newcommand{\mess}{\ensuremath{M}}

\newcommand{\neigh}{\ensuremath{\Gamma}}

\newcommand{\compat}{\ensuremath{\psi}}
\newcommand{\clipot}{\ensuremath{\phi}}

\newcommand{\eparam}{\ensuremath{\theta}}

\newcommand{\statesp}{\ensuremath{{\scr{X}}}}

\newcommand{\Part}{\ensuremath{\Phi}}

\newcommand{\indset}{\ensuremath{\scr{I}}}

\newcommand{\iset}{\ensuremath{\scr{I}}}
\newcommand{\feaset}{\ensuremath{\scr{A}}}

\newcommand{\jset}{\ensuremath{\scr{J}}}

\newcommand{\ind}{\ensuremath{i}}

\newcommand{\order}{{\mathcal{O}}}

\newcommand{\estv}{\ensuremath{\estim{\xvec}}}
\newcommand{\xmap}{\ensuremath{\estv_{\operatorname{MAP}}}}

\newcommand{\inprod}[2]{\ensuremath{\langle #1 , \, #2 \rangle}}

\newcommand{\graph}{\ensuremath{G}}
\newcommand{\subgraph}{\ensuremath{H}}
\newcommand{\vertex}{\ensuremath{V}}
\newcommand{\edge}{\ensuremath{E}}

\newcommand{\bk}{\ensuremath{\backslash}}

\newcommand{\Exs}{\ensuremath{{\mathbb{E}}}}

\newcommand{\beq}{\begin{quotation}}
\newcommand{\enq}{\end{quotation}}

\newcommand{\estart}{\begin{equation}}
\newcommand{\eend}{\end{equation}}

\newcommand{\widgraph}[2]{\includegraphics[keepaspectratio,width=#1]{#2}}

\newcommand{\scr}[1]{\ensuremath{\mathcal{#1}}}

\newcommand{\defn}{\ensuremath{:  =}}

\newcommand{\xvec}{\ensuremath{{{\mathbf{x}}}}}
\newcommand{\Xvec}{\ensuremath{{{\mathbf{X}}}}}

\newcommand{\xs}{\ensuremath{x^*}}

\newcommand{\yvec}{{\bf{y}}}

\newcommand{\bec}{\begin{center}}
\newcommand{\enc}{\end{center}}

\newcommand{\beit}{\begin{itemize}}
\newcommand{\enit}{\end{itemize}}

\newcommand{\been}{\begin{enumerate}}
\newcommand{\enen}{\end{enumerate}}

\newcommand{\comsl}{\begin{slide}}
\newcommand{\comspor}{\begin{slide*}}

\newcommand{\comsld}[2]{\begin{slide}[#1,#2]}
\newcommand{\comspord}[2]{\begin{slide*}[#1,#2]}

\newcommand{\mendsl}{\end{slide}}
\newcommand{\mendspo}{\end{slide*}}

\newcommand{\estim}[1]{\ensuremath{\widehat{#1}}}

\newcommand{\wtil}[1]{\ensuremath{\widetilde{#1}}}

\newcommand{\real}{\ensuremath{{\mathbb{R}}}}

\DeclareMathOperator{\probab}{Pr}

\DeclareMathOperator{\argmax}{argmax}

\DeclareMathOperator{\for}{for}
\DeclareMathOperator{\all}{all}
\DeclareMathOperator{\some}{some}

\newtheorem{example}{Example}
\newtheorem{lemma}{Lemma}
\newtheorem{theorem}{Theorem}
\newtheorem{corollary}{Corollary}
\newtheorem{proposition}{Proposition}

\newcommand{\cliqueord}{\ensuremath{C}}

\newcommand{\neweflatman}{\ensuremath{\mathcal{E}}}

\newcommand{\Qdual}{\ensuremath{\mathcal{Q}}}

\newcommand{\Normset}{\ensuremath{\mathcal{S}}}

\newcommand{\Fcon}{\ensuremath{\mathcal{F}}}
\newcommand{\Gcon}{\ensuremath{\mathcal{G}}}

\newcommand{\verint}{\ensuremath{\meanpar_{int}}}

\newcommand{\rub}{\ensuremath{q}}

\newcommand{\xcopt}{\ensuremath{\xvec^*}}

\newcommand{\tepvec}{\ensuremath{{\boldsymbol{\eparam}}}}
\newcommand{\tepvecs}{\ensuremath{\tepvec^*}}

\newcommand{\finex}{\ensuremath{\diamondsuit}}

\newcommand{\jcon}{\ensuremath{J}}

\newcommand{\vmean}[2]{\ensuremath{#1_{#2}}}

\newcommand{\mxvec}{\ensuremath{\xvec^*}}
\newcommand{\mx}{\ensuremath{x^*}}

\newcommand{\meontree}[2]{\ensuremath{{#1}(#2)}}

\newcommand{\contwo}{\ensuremath{C}}

\newcommand{\Myind}{\ensuremath{\delta}}

\newcommand{\pa}{\ensuremath{\pi^\treegr}}

\newcommand{\mypara}[1]{{\emph{#1}:}}

\newcommand{\algrep}{$\:$1}
\newcommand{\algtrw}{$\:$2}
\newcommand{\algtree}{$\:$3}

\comment{
\author{Martin J. Wainwright, Tommi S. Jaakkola and Alan S. Willsky}
\title{MAP estimation via agreement on trees: \\ Message-passing and
linear programming approaches}
\reportmonth{August} 
\reportyear{2003}
\reportnumber{1269}
}

\begin{document}


\title{MAP estimation via agreement on trees: Message-passing and
linear programming}

\author{M. J. Wainwright, T. S. Jaakkola and A. S. Willsky\thanks{
M. J. Wainwright ({\texttt {wainwrig@eecs.berkeley.edu}}) is with the
Department of Electrical Engineering and Computer Science and the
Department of Statistics, UC Berkeley, CA.  T. S. Jaakkola ({\texttt
{tommi@csail.mit.edu}}) and A. S. Willsky ({\texttt
{willsky@mit.edu}}) are with the Department of Electrical Engineering
and Computer Science, MIT, Cambridge, MA.  This work was presented in
part at the Allerton Conference on Communication, Computing and
Control in October 2002.  Work was supported in part by ODDR\&E MURI
Grant DAAD19-00-1-0466 through the Army Research Office; by the Office
of Naval Research through Grant N00014-00-1-0089; and by the Air Force
Office of Scientific Research through Grant F49620-00-1-0362.}}

\maketitle

\begin{center}
\vspace*{-2in}
\vbox to 2in{\footnotesize  \tt%
 \begin{tabular}[t]{c}
             Presented in part at the Allerton Conference on
Communication, Computing and Control in October 2002.\\           To
appear in the IEEE Transactions on Information Theory, November 2005.
  \end{tabular} \vfil}

\end{center}

\begin{abstract}
We develop and analyze methods for computing provably optimal {\em
maximum a posteriori} (MAP) configurations for a subclass of Markov
random fields defined on graphs with cycles.  By decomposing the
original distribution into a convex combination of tree-structured
distributions, we obtain an upper bound on the optimal value of the
original problem (i.e., the log probability of the MAP assignment) in
terms of the combined optimal values of the tree problems.  We prove
that this upper bound is tight if and only if all the tree
distributions share an optimal configuration in common.  An important
implication is that any such shared configuration must also be a MAP
configuration for the original distribution.  Next we develop two
approaches to attempting to obtain tight upper bounds: (a) a {\em
tree-relaxed linear program} (LP), which is derived from the
Lagrangian dual of the upper bounds; and (b) a {\em tree-reweighted
max-product message-passing algorithm} that is related to but distinct
from the max-product algorithm.  In this way, we establish a
connection between a certain LP relaxation of the mode-finding
problem, and a reweighted form of the max-product (min-sum)
message-passing algorithm.



\vtiny

\noindent {\bf Keywords:} Approximate inference; integer programming;
iterative decoding; linear programming relaxation; Markov random
fields; max-product algorithm; message-passing algorithms; min-sum
algorithm; MAP estimation; marginal polytope.
\end{abstract}

\normalsize

\section{Introduction}

Integer programming problems arise in various fields, including
communication theory, error-correcting coding, image processing,
statistical physics and machine learning~\cite[e.g.,]{McEliece98,
Verdu, Besag86}.  Many such problems can be formulated in terms of
Markov random fields~\cite[e.g.,]{Besag86,Cowell}, in which the cost
function corresponds to a graph-structured probability distribution,
and the goal is to find the {\em maximum a posteriori} (MAP)
configuration.  It is well-known that the complexity of solving the
MAP estimation problem on a Markov random field (MRF) depends
critically on the structure of the underlying graph.  For cycle-free
graphs (also known as trees), the MAP problem can be solved by a form
of non-serial dynamic programming known as the max-product or min-sum
algorithm~\cite[e.g.,]{Aji97,Cowell,Dawid92}. This algorithm, which entails
passing ``messages'' from node to node, represents a generalization of
the Viterbi algorithm~\cite{Viterbi67} from chains to arbitrary
cycle-free graphs.  In recent years, the max-product algorithm has
also been studied in application to graphs with cycles as a method for
computing approximate MAP
assignments~\cite[e.g.,]{Aji_maxplus98,Forney01b,Forney01c,FreemanMAP01,
Horn_phd,Wainwright02b}.  Although the method may perform well in
practice, it is no longer guaranteed to output the correct MAP
assignment, and it is straightforward to demonstrate problems on which
it specifies an incorrect (i.e., non-optimal) assignment.

\subsection{Overview}

In this paper, we present and analyze new methods for computing MAP
configurations for MRFs defined on graphs with cycles.  The basic idea
is to use a convex combination of tree-structured distributions to
derive upper bounds on the cost of a MAP configuration.  We prove that
any such bound is tight if and only if the trees share a common
optimizing configuration; moreover, any such shared configuration must
be MAP-optimal for the original problem.  Consequently, when the bound
is tight, obtaining a MAP configuration for a graphical model with
cycles --- in general, a {\em very} difficult problem --- is reduced
to the easy task of examining the optima of a collection of
tree-structured distributions.

Accordingly, we focus our attention on the problem of obtaining tight
upper bounds, and propose two methods directed to this end.  Our first
approach is based on the convexity of the upper bounds, and the
associated theory of Lagrangian duality.  We begin by re-formulating
the exact MAP estimation problem on a graph with cycles as a linear
program (LP) over the so-called marginal polytope.  We then consider
the Lagrangian dual of the problem of optimizing our upper bound.  In
particular, we prove that this dual is another LP, one which has a
natural interpretation as a relaxation of the LP for exact MAP
estimation.  The relaxation is obtained by replacing the marginal
polytope for the graph with cycles, which is a very complicated set in
general, by an outer bound with simpler structure.  This outer bound
is an exact characterization of the marginal polytope of any
tree-structured distribution, for which reason we refer to this
approach as a {\em tree-based LP relaxation}.

The second method consists of a class of message-passing algorithms
designed to find a collection of tree-structured distributions that
share a common optimum.  The resulting algorithm, though similar to
the standard max-product (or min-sum)
algorithm~\cite[e.g.,]{FreemanMAP01,Wainwright02b}, differs from it in
a number of important ways.  In particular, under the so-called
optimum specification criterion, fixed points of our {\em
tree-reweighted max-product algorithm} specify a MAP-optimal
configuration with a guarantee of correctness.  We also prove that
under this condition, fixed points of the tree-reweighted max-product
updates correspond to dual-optimal solutions of the tree-relaxed
linear program.  As a corollary, we establish that the ordinary
max-product algorithm on trees is solving the dual of an exact LP
formulation of the MAP estimation problem.  

Overall, this paper establishes connections between two approaches to
solving the MAP estimation problem: LP relaxations of integer
programming problems~\cite[e.g.,]{Bertsimas,SchrijverMon}, and
(approximate) dynamic programming methods using message-passing in the
max-product algebra.  More specifically, our work shows that a
(suitably reweighted) form of the max-product or min-sum algorithm is
very closely connected to a particular linear programming relaxation
of the MAP integer program.  This variational characterization has
links to the recent work of Yedidia et al.~\cite{Yedidia02}, who
showed that the sum-product algorithm has a variational interpretation
involving the so-called Bethe free energy.  In addition, the work
described here is linked in spirit to our previous
work~\cite{Wainwright02a,Wainwright05}, in which we showed how to
upper bound the log partition function using a ``convexified form'' of
the Bethe free energy.  Whereas this convex variational problem led to
a method for computing approximate marginal distributions, the current
paper deals exclusively with the problem of computing MAP
configurations.  Importantly and in sharp contrast with our previous
work, there is a non-trivial set of problems for which the upper
bounds of this paper are tight, in which case the MAP-optimal
configuration can be obtained by the techniques described here.

\subsection{Notes and related developments}
We briefly summarize some developments related to the ideas described
in this paper. In a parallel collaboration with Feldman and
Karger~\cite{Feldman02aller,Feldman03,Feldman05}, we have studied the
tree-relaxed linear program (LP) and related message-passing
algorithms as decoding methods for turbo-like and low-density parity
check (LDPC) codes, and provided finite-length performance guarantees
for particular codes and channels. In independent work, Koetter and
Vontobel~\cite{KoeVon03} used the notion of a graph cover to establish
connections between the ordinary max-product algorithm for LDPC codes,
and a particular polytope equivalent to the one defining our LP
relaxation.  In other independent work, Wiegerinck and
Heskes~\cite{Wiegerinck02_frac} have proposed a ``fractional'' form of
the sum-product algorithm that is closely related to the
tree-reweighted sum-product algorithm considered in our previous
work~\cite{Wainwright05}; see also Minka~\cite{Minka04} for a
reweighted version of the expectation propagation algorithm.  In other
work, Kolmogorov~\cite{Kol05,Kol05b} has studied the tree-reweighted
max-product message-passing algorithms presented here, and proposed a
sequential form of tree-updates for which certain convergence
guarantees can be established.  In follow-up work, Kolmogorov and
Wainwright~\cite{KolWai05} provided stronger optimality properties of
tree-reweighted message-passing when applied to problems with binary
variables and pairwise interactions.

\subsection{Outline} 

The remainder of this paper is organized as follows.
Section~\ref{SecBackground} provides necessary background on graph
theory and graphical models, as well as some preliminary details on
marginal polytopes, and a formulation of the MAP estimation problem.
In Section~\ref{SecUpper}, we introduce the basic form of the upper
bounds on the log probability of the MAP assignment, and then develop
necessary and sufficient conditions for these bounds to be tight.  In
Section~\ref{SecLagDual}, we first discuss how the MAP integer
programming problem has an equivalent formulation as a linear program
(LP) over the marginal polytope.  We then prove that the Lagrangian
dual of the problem of optimizing our upper bounds has a natural
interpretation as a tree-relaxation of the original LP.
Section~\ref{SecTrwMax} is devoted to the development of iterative
message-passing algorithms and their relation to the dual of the LP
relaxation.  We conclude in Section~\ref{SecDiscuss} with a discussion
and extensions to the analysis presented here.

\section{Preliminaries}

\label{SecBackground}

This section provides the background and some preliminary developments
necessary for subsequent sections.  We begin with a brief overview of
some graph-theoretic basics; we refer the reader to the
books~\cite{Biggs,Bollobas_mod} for additional background on graph
theory.  We then describe the formalism of Markov random fields; more
details can be found in various
sources~\cite[e.g.,]{Bremaud91,Cowell,Lauritzen}.  We conclude by
formulating the MAP estimation problem for a Markov random field.

\subsection{Undirected graphs}

An undirected graph $\graph = (\vertex, \edge)$ consists of a set of
nodes or vertices $\vertex =\{1, \ldots, \spnodenum\}$ that are joined
by a set of edges $\edge$.  In this paper, we consider only simple
graphs, for which multiple edges between the same pair of vertices are
forbidden.  For each $s \in \vertex$, we let $\neigh(s) = \{ \; t \in
\vertex \; | \; (s,t) \in \edge \; \}$ denote the set of {\em
neighbors} of $s$.  A {\em clique} of the graph $\graph$ is a
fully-connected subset $\clset$ of the vertex set (i.e., $(s,t) \in
\edge$ for all $s,t \in \clset$).  The clique $\clset$ is {\em
maximal} if it is not properly contained within any other clique.  A
{\em cycle} in a graph is a path from a node $s$ back to itself; that
is, a cycle consists of a sequence of distinct edges \mbox{$\{ \;
(s_0, s_1), \; (s_1, s_2), \; \ldots, (s_{k-1}, s_k) \; \}$} such that
$s_0 = s_k$.

A subgraph of $\graph = (\vertex, \edge)$ is a graph $\subgraph =
(\vertex(\subgraph), \edge(\subgraph))$ where $\vertex(\subgraph)$
(respectively $\edge(\subgraph)$) are subsets of $\vertex$
(respectively $\edge$).  Of particular importance to our analysis are
those (sub)graphs without cycles.  More precisely, a {\em tree} is a
cycle-free subgraph $\treegr = (\vertex(\treegr), \edge(\treegr))$; it
is {\em spanning} if it reaches every vertex (i.e., $\vertex(\treegr)
= \vertex$). 

\comment{ See Figure~\ref{FigTrees} for illustration of these
concepts.  \newcommand{\harg}{\hspace*{.005in}}
\begin{figure}[h]
\bec
\begin{tabular}{ccccc}
\widgraph{.28\textwidth}{graph_illus.eps} &
\harg &
\widgraph{.28\textwidth}{trnonsp_illus.eps} &
\harg &
\widgraph{.28\textwidth}{trsp_illus.eps} \\
(a) & & (b) & & (c) 
\end{tabular}
\caption{(a) Graph with cycles. (b) A tree is a cycle-free subgraph.
(c) A spanning tree reaches every vertex of the graph.}
\label{FigTrees} 
\enc
\end{figure}
}

\subsection{Markov random fields}
\label{SecExpo}
A Markov random field (MRF) is defined on the basis of an undirected
graph $\graph = (\vertex, \edge)$ in the following way.  For each $s
\in \vertex$, let $X_s$ be a random variable taking values $x_s$ in
some sample space $\statesp_s$.  This paper deals exclusively with the
discrete case, for which $X_s$ takes values in the finite alphabet
\mbox{$\statesp_s \defn \{0, \ldots, \statenum_s -1 \}$.}  By
concatenating the variables at each node, we obtain a random vector
\mbox{$\Xvec = \{ \, X_s \; | \; s \in \vertex \}$} with $\spnodenum =
|\vertex|$ elements.  Observe that $\Xvec$ itself takes values $\xvec$
in the Cartesian product space \mbox{$\stsp \defn \statesp_1 \times
\statesp_2 \times \cdots \times \statesp_\spnodenum$.}  For any subset
$\aseta \subseteq \vertex$, we let $\Xvec_\aseta$ denote the
collection $\{ X_s \; | \; s \in \aseta \}$ of random variables
associated with nodes in $\aseta$, with a similar definition for
$\xvec_\aseta$.

By the Hammersley-Clifford theorem~\cite[e.g.,]{Lauritzen}, any Markov
random fields that is strictly positive (i.e., $p(\xvec) > 0$ for all
$\xvec \in \stsp$) can defined either in terms of certain Markov
properties with respect to the graph, or in terms of a decomposition
of the distribution over cliques of the graph.  We use the latter
characterization here.  For the sake of development in the sequel, it
is convenient to describe this decomposition in exponential
form~\cite[e.g.,]{Amari01}.  We begin with some necessary notation.  A
{\em potential function} associated with a given clique $\cliqueord$
is mapping $\clipot:\stsp \rightarrow \real$ that depends only on the
subcollection $\xvec_\cliqueord \defn \{ x_s \; | \; s \in \cliqueord
\}$.  There may be a family of potential functions $\{\clipot_\sumind
\; | \; \sumind \in \iset(\cliqueord)\}$ associated with any given
clique, where $\sumind$ is an index ranging over some set
$\iset(\cliqueord)$.  Taking the union over all cliques defines the
overall index set $\iset = \cup_{\cliqueord} \iset(\cliqueord)$. The
full collection of potential functions $\{ \clipot_\sumind \; | \;
\sumind \in \iset \}$ defines a vector-valued mapping
$\clipotvec:\stsp \rightarrow \real^{\df}$, where $\df = |\iset|$ is
the total number of potential functions.  Associated with $\clipotvec$
is a real-valued vector $\eparam = \{ \; \eparam_\sumind \; | \;
\sumind \in \iset \; \}$, known as the exponential parameter vector.
For a fixed $\xvec \in \stsp$, we use
$\inprod{\eparam}{\clipotvec(\xvec)}$ to denote the ordinary Euclidean
product in $\real^{\df}$ between $\eparam$ and $\clipotvec(\xvec)$.

With this set-up, the collection of strictly positive Markov random
fields associated with the graph $\graph$ and potential functions
$\clipotvec$ can be represented as the {\em exponential family} $\{
p(\xvec; \eparam) \; | \; \eparam \in \real^\df \}$, where
\begin{eqnarray}
\label{EqnExpFam}
p(\xvec; \eparam) & \propto & \exp \big\{
\inprod{\eparam}{\clipotvec(\xvec)} \big \} \; \equiv \; \exp \big \{
\sum_{\sumind \in \iset} \eparam_\sumind \clipot_\sumind(\xvec) \big
\}.
\end{eqnarray}
Note that each vector $\eparam \in \real^\df$ indexes a particular
Markov random field $p(\xvec; \eparam)$ in this exponential family.

\begin{example}
The {\em Ising model} of statistical physics~\cite[e.g.,]{Baxter}
provides a simple illustration of a collection of MRFs in this
form. This model involves a vector $\xvec \in \{-1,1\}^\spnodenum$,
with a distribution defined by potential functions only on cliques of
size at most two (i.e., vertices and edges).  As a result, the
exponential family in this case takes the form:
\begin{eqnarray}
\label{EqnIsing} 
p(\xvec; \eparam) & \propto & \exp \big\{ \sum_{s \in \vertex} \eparam_s x_s
+ \sum_{(s,t) \in \edge} \eparam_{st} x_s x_t \big \}.
\end{eqnarray}
Here $\eparam_{st}$ is the weight on edge $(s,t)$, and $\eparam_s$ is
the parameter for node $s$.  In this case, the index set $\iset$
consists of the union $\vertex \cup \edge$.  Note that the set of
potentials $\{x_s, s \in \vertex\} \cup \{x_s x_t, (s,t) \in \edge \}$
is a basis for all multinomials on $\{-1,1\}^\spnodenum$ of maximum
degree two that respect the structure of $\graph$.
\hfill \finex
\end{example}

When the collection of potential functions $\clipotvec$ do not satisfy
any linear constraints, then the representation~\eqref{EqnExpFam} is
said to be {\em minimal}~\cite{Amari01,Barndorff78}.  For example, the
Ising model~\eqref{EqnIsing} is minimal, because there is no linear
combination of the potentials $\clipotvec = \{x_s, \; s \in \vertex\}
\cup \{x_s x_t, \; (s,t) \in \edge \}$ that is equal to a constant for
all $\xvec \in \{-1,1\}^\spnodenum$.  In contrast, it is often
convenient to consider an {\em overcomplete representation}, in which
the potential functions $\clipotvec$ do satisfy linear constraints,
and hence are no longer a basis.  More specifically, our development
in the sequel makes extensive use of an overcomplete representation in
which the basic building blocks are {\em indicator functions} of the
form $\Myind_j(x_s)$ --- the function that is equal to one if $x_s
=j$, and zero otherwise.  In particular, for a Markov random field
with interactions between at most pairs of variables, we use the
following collection of potential functions:
\begin{subequations}
\label{EqnOver}
\begin{eqnarray}
\{ \Myind_j(x_s) & \big | &  j \in \statesp_s \: \} \quad
\for \; s \; \in \; \vertex, \\
\{\Myind_j(x_s) \Myind_k(x_t) & \big | & (j,k) \in
\statesp_s \times \statesp_t \; \} \quad \for \; (s,t) \; \in \; \edge,
\end{eqnarray}
\end{subequations}
which we refer to as the {\em canonical overcomplete
representation}. This representation involves a total of $\df \defn
\sum_{s \in \vertex} \statenum_s +
\sum_{(s,t) \in \edge} \statenum_s \statenum_t$ potential functions, 
indexed by the set
\begin{multline}
\label{EqnDefnIndset}
\indset \defn \big[\cup_{s \in \vertex} \{ (s;j), j \in \statesp_s
\}\big] \; \\
\cup \; \big[\cup_{(s,t) \in \edge} \{ (st;jk), (j,k) \in
\statesp_s \times \statesp_t\} \big].
\end{multline}
The overcompleteness of the representation is manifest in various
linear constraints among the potentials; for instance, the relation
$\Myind_j(x_s) - \sum_{x_t \in \statesp_t} \Myind_j(x_s) \Myind_k(x_t) = 0$
holds for all $x_s \in \statesp_s$.  As a consequence of this
overcompleteness, there are many exponential parameters corresponding
to a given distribution (i.e., $p(\xvec; \eparam) = p(\xvec;
\eparamtil)$ for $\eparam \neq \eparamtil$).  Although this
many-to-one correspondence might seem undesirable, its usefulness is
illustrated in Section~\ref{SecTrwMax}.

The bulk of this paper focuses exclusively on MRFs with interactions
between at most pairs $(x_s, x_t)$ of random variables, which we refer
to as {\em pairwise} MRFs. In principle, there is no loss of
generality in restricting to pairwise interactions, since any factor
graph over discrete variables can be converted to this form by
introducing auxiliary random variables~\cite{FreemanMAP01}; see
Appendix~\ref{AppConv} for the details of this procedure.  Moreover,
the techniques described in this paper can all be generalized to apply
directly to MRFs that involve higher-order interactions, by dealing
with hypertrees as opposed to ordinary trees.\footnote{For brevity, we
do not discuss hypertrees at length in this paper.  Roughly speaking,
they amount to trees formed on clusters of nodes from the original
graph; see Wainwright et al.~\cite{Wainwright01a} for further details
on hypertrees.}  Moreover, with the exception of specific examples
involving the Ising model, we exclusively use the canonical
overcomplete representation~\eqref{EqnOver} defined in terms of
indicator functions.

\subsection{Marginal distributions on graphs}
\label{SecMargpoly}

Our analysis in the sequel focuses on the local marginal distributions
that are defined by the indicator functions in the canonical
overcomplete representation~\eqref{EqnOver}.  In particular, taking
expectations of these indicators with respect to some distribution
$p(\cdot)$ yields marginal probabilities for each node $s \in \vertex$
\begin{eqnarray}
\label{EqnDefnMarga}
\meanpar_{s;j} & = & \Exs_p[\Myind_j(x_s)] \; \defn \; 
\sum_{\xvec \in \stsp} p(\xvec) \Myind_j(x_s)
\end{eqnarray}
and for each edge $(s,t) \in \edge$
\begin{eqnarray}
\label{EqnDefnMargb}
\meanpar_{st;jk} & \defn & \Exs_p[\Myind_j(x_s) \Myind_k(x_t)] \\
& = & \sum_{\xvec \in \stsp} p(\xvec) \; [\Myind_j(x_s)
\Myind_k(x_t)].
\end{eqnarray}
Note that equations~\eqref{EqnDefnMarga} and~\eqref{EqnDefnMargb}
define a $\df$-dimensional vector $\meanpar = \{\meanpar_\sumind,
\sumind \in \iset \}$ of marginals, indexed by elements of $\indset$
defined in equation~\eqref{EqnDefnIndset}.  We let $\Margset(\graph)$
denote the set of all such marginals realizable in this way:
\begin{multline}
\Margset(\graph)\; \defn \; \{ \meanpar \in \real^\df \; | \;
\meanpar_{s;j} = \Exs_p[\Myind_j(x_s)], \quad \mbox{and} \\
\meanpar_{st;jk} = \Exs_p[\Myind_j(x_s) \Myind_k(x_t)] \; \; \for \;
\; \some \; \; p(\cdot) \; \}.
\end{multline}
The conditions defining membership in $\Margset(\graph)$ can be
expressed more compactly in the equivalent vector form \mbox{$\meanpar
= \Exs_p[\clipotvec(\xvec)] = \sum_{\xvec \in \stsp} p(\xvec)
\clipotvec(\xvec)$,} where $\clipotvec$ denotes a vector consisting of
the potential functions forming the canonical overcomplete
representation~\eqref{EqnOver}.  We refer to $\Margset(\graph)$ as the
{\em marginal polytope} associated with the graph $\graph$.

By definition, any marginal polytope is the convex hull of a finite
number of vectors --- namely, the collection $\{ \clipotvec(\xvec) \;
| \; \xvec \in \stsp \}$.  Consequently, the Minkowski-Weyl
theorem~\cite{Rockafellar} ensures that $\Margset(\graph)$ can be
represented as an intersection of half-spaces $\cap_{j \in \jset}
H_{a_j, b_j}$ where $\jset$ is a {\em finite} index set and each
half-space is of the form $H_{a_j, b_j} \defn \{ \meanpar \in
\real^\df \; | \; \inprod{a_j}{\meanpar} \leq b_j \}$ for some $a_j
\in \real^\df$, and $b_j \in \real$.  These half-space constraints
include the non-negativity condition $\meanpar_\sumind \geq 0$ for
each $\sumind \in \iset$.  Moreover, due to the overcompleteness of
the canonical overcomplete representation, there are various
equality\footnote{Any equality constraint $\inprod{a}{\meanpar} = b$
is equivalent to enforcing the pair of inequality constraints
$\inprod{a}{\meanpar} \leq b$ and $\inprod{-a}{\meanpar} \leq -b$.}
constraints that must hold; for instance, for all nodes $s \in
\vertex$, we have the constraint $\sum_{j \in \statesp_s}
\meanpar_{s;j} = 1$ .

The number of additional (non-trivial) linear constraints required to
characterize $\Margset(\graph)$, though always finite, grows rapidly
in $\spnodenum$ for a general graph with cycles; see Deza and
Laurent~\cite{Deza97} for discussion of the binary case.  It is
straightforward, however, to specify a {\em subset} of constraints
that any $\meanpar \in \Margset(\graph)$ must satisfy.  First, as
mentioned previously, since the elements of $\meanpar$ are marginal
probabilities, we must have $\meanpar \geq 0$ (meaning that $\meanpar$
is in the positive orthant).  Second, as local marginals, the elements
of $\meanpar$ must satisfy the {\em normalization constraints}:
\begin{subequations}
\begin{eqnarray}
\label{EqnDefnNormalizea}
\sum_{j \in \statesp_s} \meanpar_{s;j} & = &  1 \quad \forall \; s \in
\vertex, \\
\label{EqnDefnNormalizeb}
\sum_{(j,k) \in \statesp_s \times \statesp_t } \meanpar_{st;jk} & = &
1 \quad \forall \; (s,t) \in \edge.
\end{eqnarray}
\end{subequations}
Third, since the single node marginal over $x_s$ must be consistent
with the joint marginal on $(x_s, x_t)$, the following {\em
marginalization constraint} must also be satisfied:
\begin{eqnarray}
\label{EqnDefnMarginalize}
\sum_{k \in \statesp_t} \meanpar_{st;jk} & = & \meanpar_{s;j}  \quad
\forall \;  (s,t) \in \edge, \; j \in \statesp_s.
\end{eqnarray}
On the basis of these constraints,\footnote{Note that the
normalization constraint on $\{\meanpar_{st;jk} \}$ is redundant given
the marginalization constraint~\eqref{EqnDefnMarginalize}, and the
normalization of $\{\meanpar_{s;j} \}$.}  we define the set
$\Locset(\graph)$ as all $\meanpar \in \real_+^\df$ that satisfy
constraints~\eqref{EqnDefnNormalizea}, \eqref{EqnDefnNormalizeb},
and~\eqref{EqnDefnMarginalize}.  Here it should be understood that
there are two sets of marginalization constraints for each edge
$(s,t)$: one for each of the variables $x_s$ and $x_t$.  By
construction, $\Locset(\graph)$ specifies an outer bound on
$\Margset(\graph)$; moreover, in contrast to $\Margset(\graph)$, it
involves only a number of inequalities that is polynomial in
$\spnodenum$. More specifically, $\Locset(\graph)$ is defined by
$\order(\statenum \spnodenum + \statenum^2 |\edge|)$ inequalities,
where $\statenum \defn \max_{s} |\statesp_s|$.  Since the number of
edges $|\edge|$ is at most ${\spnodenum \choose 2}$, this complexity
is at most $\order(\statenum^2 \spnodenum^2)$.  The constraint set
$\Locset(\graph)$ plays an important role in the sequel.

\subsection{MAP estimation}

Of central interest in this paper is the computation of {\em maximum a
posteriori} (MAP) configurations\footnote{The term a posteriori arises
from applications, in which case one often wants to compute maximizing
elements of the posterior distribution $p(\xvec \, | \yvec; \eparam)$,
where $\yvec$ is a fixed collection of noisy observations.} for a
given distribution in an exponential form --- i.e., configurations in
the set $\arg \max_{\xvec \in \stsp} p(\xvec; \eparambar)$, where
$\eparambar \in \real^\df$ is a given vector of weights.  For reasons
to be clarified, we refer to $p(\xvec; \eparambar)$ as the {\em target
distribution}.  The problem of computing a MAP configuration arises in
a wide variety of applications.  For example, in image
processing~\cite[e.g.,]{Besag86}, computing MAP estimates can be used
as the basis for image segmentation techniques.  In error-correcting
coding~\cite[e.g.,]{McEliece98}, a decoder based on computing the MAP
codeword minimizes the word error rate.

When using the canonical overcomplete representation
$\clipotvec(\xvec) = \{\Myind_j(x_s), \; \Myind_j(x_s) \Myind_k(x_t)
\}$, it is often convenient to represent the exponential parameters in
the following functional form:
\begin{subequations}
\label{EqnFuncNote}
\begin{eqnarray}
\label{EqnFuncNoteA}
\eparambar_s(x_s) & \defn & \sum_{j \in \statesp_s} \eparambar_{s;j}
\Myind_j(x_s), \\
\label{EqnFuncNoteB}
\eparambar_{st}(x_s, x_t) &  \defn & \sum_{(j,k) \in \statesp_s
\times \statesp_t} \eparambar_{st;jk} \Myind_j(x_s) \Myind_k(x_t).
\end{eqnarray}
\end{subequations}
With this notation, the MAP problem is equivalent to finding a
configuration $\xmap \in \stsp$ that maximizes the quantity
\begin{eqnarray}
\label{EqnMAPProblem}
 \inprod{\eparambar}{\clipotvec(\xvec)} & \defn & \sum_{s \in \vertex}
\eparambar_s(x_s) + \sum_{(s,t) \in \edge} \eparambar_{st}(x_s, x_t).
\end{eqnarray}

Although the parameter $\eparambar$ is a known and fixed quantity, it
is useful for analytical purposes to view it as a variable, and define
a function $\Partinf(\eparambar)$ as follows:
\begin{eqnarray}
\label{EqnDefnPartinf}
\Partinf(\eparambar) & \defn & \max_{\xvec \in \stsp}
\inprod{\eparambar}{\clipotvec(\xvec)}.
\end{eqnarray}
Note that $\Partinf(\eparambar)$ represents the value of the optimal
(MAP) configuration as $\eparambar$ ranges over $\real^\df$.  As the
maximum of a collection of linear functions, $\Partinf$ is convex in
terms of $\eparambar$.

\section{Upper bounds via convex combinations}
\label{SecUpper}

This section introduces the basic form of the upper bounds on
$\Partinf(\eparambar)$ to be considered in this paper.  The key
property of $\Partinf$ is its convexity, which allows us to apply
Jensen's inequality~\cite{Hiriart1}.  More specifically, let
$\{\treedist^i\}$ be a finite collection of non-negative weights that
sum to one, and consider a collection $\{\eparam^i\}$ of exponential
parameters such that $\sum_{i} \treedist^i \eparam^i = \eparambar$.
Then applying Jensen's inequality yields the upper bound
\begin{eqnarray}
\label{EqnBasicJen}
\Partinf(\eparambar) & \leq & \sum_{i} \treedist^i
\Partinf(\eparam^i).
\end{eqnarray}
Note that the bound~\eqref{EqnBasicJen} holds for any collection of
exponential parameters $\{\eparam^i\}$ that satisfy $\sum_i
\treedist^i \eparam^i = \eparambar$; however, the bound will not
necessarily be useful, unless the evaluation of $\Partinf(\eparam^i)$
is easier than the original problem of computing
$\Partinf(\eparambar)$.  Accordingly, in this paper, we focus on
convex combinations of {\em tree-structured} exponential parameters
(i.e., the set of non-zero components of $\eparam^i$ is restricted to
an acyclic subgraph of the full graph), for which exact computations
are tractable.  In this case, each index $i$ in
equation~\eqref{EqnBasicJen} corresponds to a spanning tree of the
graph, and the corresponding exponential parameter is required to
respect the structure of the tree.  In the following, we introduce the
necessary notation required to make this idea precise.

\subsection{Convex combinations of trees}
\label{SecConvexTree}
For a given graph, let $\treegr$ denote a particular spanning tree,
and let $\tract = \tract(\graph)$ denote the set of all spanning
trees.  For a given spanning tree $\treegr = (\vertex,
\edge(\treegr))$, we define a set
\begin{multline*}
\iset(\treegr) = \{ (s;j) \; | \; s \in \vertex, \; j \in \statesp_s
\} \; \; \\
\cup \; \; \{ (st; jk) \; | \; (s,t) \in \edge(\treegr), \; (j,k) \in
\statesp_s \times \statesp_t \},
\end{multline*}
corresponding to those indexes associated with all vertices but only
edges in the tree.

To each spanning tree $\treegr \in \tract$, we associate an
exponential parameter $\eparam(\treegr)$ that must respect the
structure of $\treegr$.  More precisely, the parameter
$\eparam(\treegr)$ must belong to the linear constraint set
$\neweflatman(\treegr)$ given by
\begin{equation}
\label{EqnEflat}
 \{ \; \eparam(\treegr) \in \real^\df \; | \; \eparam_\sumind(\treegr)
= 0 \; \; \forall \; \sumind \in \iset\bk \iset(\treegr) \; \}.
\end{equation}
By concatenating all of the tree vectors, we form a larger vector
$\tepvec = \{ \eparam(\treegr), \; \treegr \in \tract \}$, which is an
element of $\real^{\df \times |\tract(\graph)|}$.  The vector
$\tepvec$ must belong to the constraint set
\begin{equation}
\label{EqnDefnNeweflatman}
\neweflatman \defn  \{ \tepvec \in \real^{\df \times
|\tract(\graph)|} \; \big | \; \eparam(\treegr) \in \neweflatman(\treegr)
\; \; \forall \; \treegr \; \in \; \tract(\graph) \}.
\end{equation}

\newcommand{\hspew}{\hspace*{.4in}}
\newcommand{\spfigsize}{0.076\textwidth}

\begin{figure}[h]
\bec
\begin{tabular}{ccc}
\widgraph{\spfigsize}{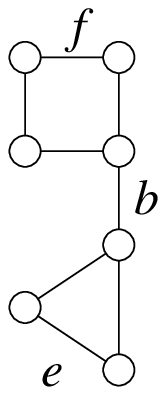} &
\hspew &
\widgraph{\spfigsize}{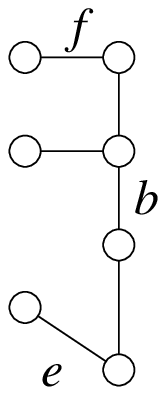} \\
(a) & & (b) \\
& & \\
\widgraph{\spfigsize}{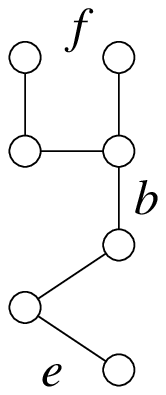} & 
\hspew &
\widgraph{\spfigsize}{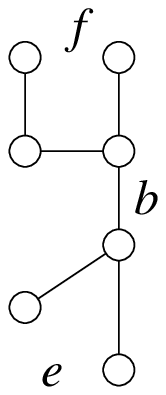} \\
(c) &  & (d)
\end{tabular}
\caption{Illustration of edge appearance probabilities.  Original
graph is shown in panel (a).  Probability $1/3$ is assigned to each of
the three spanning trees $\{ \, \treegr_\tind \; | \; \tind = 1,2,3 \,
\}$ shown in panels (b)--(d). Edge $b$ is a so-called bridge in
$\graph$, meaning that it must appear in any spanning tree. Therefore,
it has edge appearance probability $\treedist_b = 1$.  Edges $e$ and
$f$ appear in two and one of the spanning trees, respectively, which
gives rise to edge appearance probabilities $\treedist_e = 2/3$ and
$\treedist_f = 1/3$.}
\label{FigEdgeAppear}
\enc
\end{figure}

In order to define convex combinations of exponential parameters
defined on spanning trees, we require a probability distribution
$\treedistvec$ over the set of spanning trees
\begin{equation*}
\treedistvec \defn \{ \treedist(\treegr) \; \big |
\; \treedist(\treegr) \geq 0, \; \; \; \sum_{\treegr \in \tract}
\treedist(\treegr) = 1 \; \}.
\end{equation*}
For any distribution $\treedistvec$, we define its {\em support} to be the
set of trees to which it assigns strictly positive probability; that
is
\begin{eqnarray}
\label{EqnDefnSupp}
\supp(\treedistvec) & \defn & \{ \; \treegr \in \tract \; | \;
\treedist(\treegr) > 0 \; \}.
\end{eqnarray}
In the sequel, it will also be of interest to consider the probability
$\treedist_e = \probab_\treedistvec \{ e \in \treegr \}$ that a given
edge $e \in \edge$ appears in a spanning tree $\treegr$ chosen
randomly under $\treedistvec$.  We let $\trdistvec = \{ \treedist_e \;
| \; e \in \edge \}$ represent a vector of these {\em edge appearance
probabilities}.  Any such vector $\trdistvec$ must belong to the
so-called {\em spanning tree polytope}~\cite{Bertsimas,Edmonds71},
which we denote by $\treepoly(\graph)$.  See
Figure~\ref{FigEdgeAppear} for an illustration of the edge appearance
probabilities.  Although we allow for the support
$\supp(\treedistvec)$ to be a strict subset of the set of all spanning
trees, we require that $\treedist_e > 0$ for all $e
\in \edge$, so that each edge appears in at least one tree with
non-zero probability.

Given a collection of tree-structured parameters $\tepvec$ and a
distribution $\treedistvec$, we form a {\em convex combination} of
tree exponential parameters as follows
\begin{eqnarray}
\label{EqnConvComb}
\Exs_{\treedistvec}[\eparam(\treegr)] & \defn \sum \limits_{\treegr}
\treedist(\treegr) \eparam(\treegr).
\end{eqnarray}
Let $\eparambar \in \real^\df$ be the target parameter vector for
which we are interested in computing $\Partinf$, as well as a MAP
configuration of $p(\xvec; \eparambar)$.  For a given $\treedistvec$,
of interest are collections $\tepvec$ of tree-structured exponential
parameters such that $\Exs_{\treedistvec}[\eparam(\treegr)] =
\eparambar$.  Accordingly, we define the following constraint set:
\begin{eqnarray}
\label{EqnDefnFeaset}
\feaset_\treedistvec(\eparambar) & \defn & \{ \tepvec \in \neweflatman \; | \;
\Exs_{\treedistvec}[\eparam(\treegr)] = \eparambar \}.
\end{eqnarray}
It can be seen that $\feaset_\treedistvec(\eparambar)$ is never empty
as long as $\treedist_e > 0$ for all edges $e \in \edge$.  We say that
any member $\tepvec$ of $\feaset_\treedistvec(\eparambar)$ specifies a
{\em $\treedistvec$-reparameterization} of $p(\xvec; \eparambar)$.

\begin{example}[Single cycle]
\label{ExaSingLoop1}
To illustrate these definitions, consider a binary vector $\xvec \in
\{0,1\}^4$ on a 4-node cycle, with the distribution in the minimal
Ising form 
\begin{eqnarray*}
p(\xvec; \eparambar) & \propto & \exp \{x_1 x_2 + x_2 x_3 + x_3 x_4 +
x_4 x_1 \}.
\end{eqnarray*}
In words, the target distribution is specified by the minimal
parameter $\eparambar \ = [0 \; 0 \; 0 \; 0 \: \; 1 \; 1 \; 1 \; 1]$,
where the zeros represent the fact that $\eparambar_s = 0$ for all $s
\in \vertex$.
\comment{
\begin{figure*}[t]
\bec
\begin{tabular}{cccccc}
\widgraph{.17\textwidth}{fin_loopcut1.eps} &
\hspace*{.08in} &
\widgraph{.17\textwidth}{fin_loopcut2.eps} 
&
\widgraph{.17\textwidth}{fin_loopcut3.eps} & 
\hspace*{.08in} & 
\widgraph{.17\textwidth}{fin_loopcut4.eps}
\end{tabular}
\enc
\caption[Convex combination of distributions]{A convex combination of
four distributions $p(\xvec; \eparam(\treegr_\ind))$, each defined by a
spanning tree $\treegr_\ind$, is used to approximate the target
distribution $p(\xvec; \eparambar)$ on the single-cycle graph.}
\label{FigSingLoop}
\end{figure*}
} Suppose that $\treedistvec$ is the uniform distribution
$\treedist(\treegr_\ind) = 1/4$ for $\ind=1, \ldots 4$, so that
$\treedist_e = 3/4$ for each edge $e \in \edge$.  We construct a
member $\tepvec$ of $\feaset_\treedistvec(\eparambar)$, as follows:
\begin{eqnarray*}
\eparam(\treegr_1) & = & (4/3) \; \begin{bmatrix} 0 & 0 & 0 & 0 & 1 &
1 & 1 & 0
\end{bmatrix}, \\
\eparam(\treegr_2) & = & (4/3) \; \begin{bmatrix} 0 & 0 & 0 & 0 & 1 &
1 & 0 & 1
\end{bmatrix},  \\
\eparam(\treegr_3) & = & (4/3) \; \begin{bmatrix} 0 & 0 & 0 & 0 & 1 &
0 & 1 & 1
\end{bmatrix}, \\
\eparam(\treegr_4) & = & (4/3) \; \begin{bmatrix} 0 & 0 & 0 & 0 & 0 &
1 & 1 & 1
\end{bmatrix}. 
\end{eqnarray*}
With this choice, it is easily verified that
$\Exs_{\treedistvec}[\eparam(\treegr)] = \eparambar$ so that
$\tepvec \in \feaset_\treedistvec(\eparambar)$.
\end{example}

\subsection{Tightness of upper bounds}
\label{SecBasicBou}

It follows from equations~\eqref{EqnBasicJen},~\eqref{EqnConvComb}
and~\eqref{EqnDefnFeaset} that for any \mbox{$\tepvec \in
\feaset_\treedistvec(\eparambar)$,} there holds:
\begin{eqnarray}
\label{EqnBasicUpper}
\Partinf(\eparambar) & \leq & \sum_{\treegr} \treedist(\treegr)
\Partinf(\eparam(\treegr)) \nonumber \\
& = & \sum_{\treegr} \treedist(\treegr) \max_{\xvec \in \stsp} \big \{
\inprod{\eparam(\treegr)}{\clipotvec(\xvec)} \big \}.
\end{eqnarray}
Our first goal is to understand when the upper
bound~\eqref{EqnBasicUpper} is {\em tight} --- that is, met with
equality.  It turns out that that equality holds if and only if the
collection of trees share a common optimum, which leads to the notion
of {\em tree agreement}.

More formally, for any exponential parameter vector $\eparam \in
\real^\df$, define the collection $\xoptset(\eparam)$ of its optimal
configurations as follows:
\begin{equation}
 \{ \xvec \in \stsp \; | \; \inprod{\eparam}{\clipotvec(\xvec')} \;
\leq \inprod{\eparam}{\clipotvec(\xvec)} \; \; \for \; \all \; \xvec'
\in \stsp \}.
\end{equation}.
Note that by the definition~\eqref{EqnDefnPartinf} of $\Partinf$,
there holds $\inprod{\eparambar}{\clipotvec(\xvec)} =
\Partinf(\eparambar)$ for any $\xvec \in \xoptset(\eparambar)$.  With
this notation, we have:
\begin{proposition}[Tree agreement]
\label{PropTightBound}
Let $\tepvec = \{\eparam(\treegr)\} \in
\feaset_\treedistvec(\eparambar)$, and let $\cap_{\treegr \in
\supp(\treedistvec)} \xoptset(\eparam(\treegr))$ be the set of
configurations that are optimal for every tree-structured
distribution.  Then the following containment always holds:
\begin{eqnarray}
\cap_{\treegr \in \supp(\treedistvec)}
\xoptset(\eparam(\treegr))  & \subseteq & \xoptset(\eparambar).
\end{eqnarray}
Moreover, the bound~\eqref{EqnBasicUpper} is tight if and only if the
intersection on the LHS is non-empty.
\end{proposition}
\begin{proof}
The containment relation is clear from the form of the upper
bound~\eqref{EqnBasicUpper}. Let $\xcopt$ belong to
$\xoptset(\eparambar)$.  Then the difference of the RHS and the LHS of
equation~\eqref{EqnBasicUpper} can be written as follows:
\begin{eqnarray*}
0 & \leq & \Big [\sum_{\treegr} \treedist(\treegr)
\Partinf(\eparam(\treegr)) \Big ] - \Partinf(\eparambar) \\
& = & \Big [\sum_{\treegr} \treedist(\treegr)
\Partinf(\eparam(\treegr)) \Big ] -
\inprod{\eparambar}{\clipotvec(\xcopt)} \nonumber \\
& = & \sum_{\treegr} \treedist(\treegr) \big [
\Partinf(\eparam(\treegr)) -
\inprod{\eparam(\treegr)}{\clipotvec(\xcopt)} \big ], \nonumber
\end{eqnarray*}
where the last equality uses the fact that $\sum_{\treegr}
\treedist(\treegr) \eparam(\treegr) = \eparambar$.  Now for each
$\treegr \in \supp(\treedistvec)$, the term
$\Partinf(\eparam(\treegr)) -
\inprod{\eparam(\treegr)}{\clipotvec(\xcopt)}$ is non-negative, and
equal to zero only when $\xcopt$ belongs to
$\xoptset(\eparam(\treegr))$.  Therefore, the bound is tight if and
only if $\xcopt \in \cap_{\treegr \in \supp(\treedistvec)}
\xoptset(\eparam(\treegr))$ for some $\xcopt \in
\xoptset(\eparambar)$.
\end{proof} \hfill  \\

The preceding result shows that the upper bound~\eqref{EqnBasicUpper}
is tight if and only if all the trees in the support of $\treedistvec$
agree on a common configuration.  When this {\em tree agreement}
condition holds, a MAP configuration for the original problem
$p(\xvec; \eparambar)$ can be obtained simply by examining the
intersection $\cap_{\treegr \in
\supp(\treedistvec)}
\xoptset(\eparam(\treegr))$ of configurations that are optimal on
every tree for which \mbox{$\treedist(\treegr) > 0$.}  Accordingly, we
focus our attention on the problem of finding upper
bounds~\eqref{EqnBasicUpper} that are tight, so that a MAP
configuration can be obtained.  Since the target parameter
$\eparambar$ is fixed and assuming that we fix the spanning tree
distribution $\treedistvec$, the problem on which we focus is that of
optimizing the upper bound as a function of $\tepvec \in
\feaset_\treedistvec(\eparambar)$.  Proposition~\ref{PropTightBound}
suggests two different strategies for attempting to find a tight upper
bound, which are the subjects of the next two sections:  \\

\mypara{Direct minimization and Lagrangian duality}
The first approach is a direct one, based on minimizing
equation~\eqref{EqnBasicUpper}.  In particular, for a fixed
distribution $\treedistvec$ over spanning trees, we consider the
constrained optimization problem of minimizing the RHS of
equation~\eqref{EqnBasicUpper} subject to the constraint $\tepvec \in
\feaset_\treedistvec(\eparambar)$.  The problem structure ensures that
strong duality holds, so that it can be tackled via its Lagrangian
dual.  In Section~\ref{SecLagDual}, we show that this dual problem is
a linear programming (LP) relaxation of the original MAP estimation
problem.

\vtiny

\mypara{Message-passing approach}
In Section~\ref{SecTrwMax}, we derive and analyze message-passing
algorithms, the goal of which is to find, for a fixed distribution
$\treedistvec$, a collection of exponential parameters $\tepvecs = \{
\eparams(\treegr) \}$ such that $\tepvecs$ belongs to the constraint
set $\feaset_\treedistvec(\eparambar)$ of
equation~\eqref{EqnDefnFeaset}, and the intersection $\cap_\treegr
\xoptset(\eparams(\treegr))$ of configurations optimal for all tree
problems is non-empty.  Under these conditions,
Proposition~\ref{PropTightBound} guarantees that for all
configurations in the intersection, the bound is tight.  In
Section~\ref{SecTrwMax}, we develop a class of message-passing
algorithms with these two goals in mind.  We also prove that when the
bound is tight, fixed points of these algorithms specify optimal
solutions to the LP relaxation derived in Section~\ref{SecLagDual}.

\section{Lagrangian duality and tree relaxation}
\label{SecLagDual}
In this section, we develop and analyze a Lagrangian reformulation of
the problem of optimizing the upper bounds ---- i.e., minimizing the
RHS of equation~\eqref{EqnBasicUpper} as a function of $\tepvec \in
\feaset_\treedistvec(\eparambar)$.  The cost function is a linear
combination of convex functions, and so is also convex as a function
of $\tepvec$; moreover, the constraints are linear in $\tepvec$.
Therefore, the minimization problem can be solved via its Lagrangian
dual.  Before deriving this dual, it is convenient to develop an
alternative representation of $\Partinf$ as a linear program.

\subsection{Linear program over the marginal polytope for exact MAP estimation} 

Recall from equation~\eqref{EqnDefnPartinf} that the function value
$\Partinf(\eparambar)$ corresponds to the optimal value of the integer
program \mbox{$\max_{\xvec} \inprod{\eparambar}{\clipotvec(\xvec)}$.}
We now reformulate this integer program as a linear program (LP),
which leads to an alternative representation of the function
$\Partinf$, and hence of the exact MAP estimation problem. In order to
convert from integer to linear program, our approach is the standard
one~\cite[e.g.,]{Bertsimas,SchrijverMon} of taking convex combinations of all
possible solutions.  The resulting convex hull is precisely the
marginal polytope $\Margset(\graph)$ defined in
Section~\ref{SecMargpoly}.  We summarize in the following:
\begin{lemma}
\label{LemMargRep}
The function $\Partinf(\eparambar)$ has an alternative representation
as a linear program over the marginal polytope:
\begin{eqnarray}
\label{EqnMargRep}
\Partinf(\eparambar) & = & \max_{\meanparb \in \Margset(\graph) }
\inprod{\eparambar}{\meanpar}
\end{eqnarray}
where $\inprod{\eparambar}{\meanpar}$ is shorthand for the sum
$\sum_{s \in \vertex} \sum_j \meanpar_{s;j} \eparambar_{s;j} +
\sum_{(s,t) \in \edge}  \sum_{j,k} \meanpar_{st;jk} \eparambar_{st;jk}$.
\end{lemma}
\begin{proof} 
Although this type of LP reformulation is standard in combinatorial
optimization, we provide a proof here for completeness. Consider the
set $\mathcal{P} \defn \{\; p(\cdot) \; | \; p(\xvec) \geq 0, \; \;
\sum_{\xvec} p(\xvec) = 1 \}$ of all possible probability
distributions over $\xvec$.  We first claim that the maximization over
$\xvec \in \stsp$ can be rewritten as an equivalent maximization over
$\mathcal{P}$ as follows:
\begin{eqnarray}
\label{EqnEqual}
\max_{\xvec \in \stsp} \inprod{\eparambar}{\clipotvec(\xvec)} & = &
\max_{p \in \mathcal{P}} \biggr \{ \sum_{\xvec \in \stsp} p(\xvec)
\inprod{\eparambar}{\clipotvec(\xvec)} \biggr \}.
\end{eqnarray}
On one hand, the RHS is certainly greater than or equal to the LHS,
because for any configuration $\xvec^*$, the set $\mathcal{P}$
includes the delta distribution that places all its mass at $\xvec^*$.
On the other hand, for any $p \in \mathcal{P}$, the sum $\sum_{\xvec
\in \stsp} p(\xvec) \inprod{\eparambar}{\clipotvec(\xvec)}$ is a
convex combination of terms of the form
$\inprod{\eparambar}{\clipotvec(\xvec)}$ for $\xvec \in \stsp$, and so
cannot be any larger than $\max_{\xvec}
\inprod{\eparambar}{\clipotvec(\xvec)}$.

Making use of the functional notation in equation~\eqref{EqnFuncNote},
we now expand the summation on the RHS of equation~\eqref{EqnEqual},
and then use the linearity of expectation to write:
\begin{multline*}
\sum_{\xvec \in \stsp} p(\xvec) \biggr \{\sum_{s \in \vertex}
\eparambar_s(x_s) + \sum_{(s,t) \in \edge} \eparambar_{st}(x_s, x_t)
\biggr \}  =  \\
 \sum_{s \in \vertex} \sum_{j \in \statesp_s} \eparambar_{s;j}
\meanpar_{s;j} + \sum_{{\fn {(s,t) \in \edge}}} \; \; \sum_{(j,k) \in
\statesp_s \times \statesp_t} \eparambar_{st;jk} \meanpar_{st;jk}.
\end{multline*}
Here $\meanpar_{s;j} \defn \sum_{\xvec \in \stsp} p(\xvec)
\Myind_{j}(x_s)$ and $\meanpar_{st;jk} \defn \sum_{\xvec \in \stsp}
p(\xvec) \Myind_{jk} (x_s, x_t)$.  As $p$ ranges over $\mathcal{P}$, the
marginals $\meanpar$ range over $\Margset(\graph)$.  Therefore, we
conclude that $\max_{\xvec \in \stsp}
\inprod{\eparambar}{\clipotvec(\xvec)}$ is equal to $\max_{\meanpar
\in \Margset(\graph)} \inprod{\eparambar}{\meanpar}$, as claimed.
\end{proof} \hfill  \\

\mypara{Remarks}    
Lemma~\ref{LemMargRep} identifies $\Partinf(\eparambar)$ as the {\em
support function}~\cite{Hiriart1} of the set $\Margset(\graph)$.
Consequently, $\Partinf(\eparambar)$ can be interpreted as the
negative intercept of the supporting hyperplane to $\Margset(\graph)$
with normal vector $\eparambar \in \real^\df$.  This property
underlies the dual relation that is the focus of the following
section.

\subsection{Lagrangian dual}
\label{SecSublag}

Let us now address the problem of finding the tightest upper bound of
the form in equation~\eqref{EqnBasicUpper}.  More formally, for a
fixed distribution $\treedistvec$ over spanning trees, we wish to
solve the constrained optimization problem:
\begin{equation}
\label{EqnBasicProb}
\begin{cases} 
\min \limits_{\tepvec \in \neweflatman} & \sum_{\treegr}
\treedist(\treegr) \Partinf(\eparam(\treegr)) \\
\suchthat & \sum_{\treegr} \treedist(\treegr) \eparam(\treegr) =
\eparambar. 
\end{cases} 
\end{equation}
As defined in equation~\eqref{EqnDefnNeweflatman}, the constraint set
$\neweflatman$ consists of all vectors $\tepvec = \{\eparam(\treegr)
\}$ such that for each tree $\treegr$, the subvector
$\eparam(\treegr)$ respects the structure of $\treegr$, meaning that
$\eparam_\sumind(\treegr) = 0 \; \; \forall \; \sumind \in \iset\bk
\iset(\treegr)$.

Note that the cost function is a convex combination of convex
functions; moreover, with $\treedistvec$ fixed, the constraints are
all linear in $\tepvec$.  Under these conditions, strong duality
holds~\cite{Bertsekas_nonlin}, so that this constrained optimization
problem can be tackled via its Lagrangian dual.  The dual formulation
turns out to have a surprisingly simple and intuitive form.  In
particular, recall the set $\Locset(\graph)$ defined the orthant
constraint $\taupar \in \real_+^\df$, and the additional linear
constraints~\eqref{EqnDefnNormalizea}, \eqref{EqnDefnNormalizeb}
and~\eqref{EqnDefnMarginalize}.  The polytope $\Locset(\graph)$ turns out
to be the constraint set in the dual reformulation of our problem:
\begin{theorem}
\label{ThmLagDual}
The Lagrangian dual to problem~\eqref{EqnBasicProb} is given by the LP
relaxation based on $\Locset(\graph)$.  Given that strong duality
holds, the optimal primal value 
\begin{equation}
\label{EqnTreeRelaxA}
\min \limits_{\tepvec \in \neweflatman \; \mbox{and} \; \sum_{\treegr}
\treedist(\treegr) \eparam(\treegr) \treedist(\treegr) }
\Partinf(\eparam(\treegr)) 
\end{equation}
is equal to the optimal dual value
\begin{equation}
\label{EqnTreeRelaxB}
 \max_{\tausymb \in \Locset(\graph)} \big \{ \sum_{s \in \vertex}
 \sum_{j} \tausym_{s;j} \eparambar_{s;j} + \sum_{(s,t) \in \edge} \;
 \; \; \sum_{(j,k)} \tausym_{st;jk} \eparambar_{st;jk} \big \}.
\end{equation}
\end{theorem}
\begin{proof}
Let $\tausymb$ be a vector of Lagrange multipliers corresponding to
the constraints $\Exs_{\treedistvec}[\eparam(\treegr)] = \eparambar$.
We then form the Lagrangian associated with
problem~\eqref{EqnBasicProb}:
\begin{eqnarray*}
\Lag_{\treedistvec, \eparambar}(\tepvec, \tausymb) & = &
\Exs_{\treedistvec}[ \Partinf(\eparam(\treegr))] +
\inprod{\tausymb}{\eparambar - \sum_{\treegr} \treedist(\treegr)
\eparam(\treegr)} \nonumber \\
\label{EqnLag}
& = & \sum_{\treegr} \treedist(\treegr) \; \big [
\Partinf(\eparam(\treegr)) - \inprod{\eparam(\treegr)}{\tausymb} \big]
\; + \; \inprod{\tausymb}{\eparambar}.
\end{eqnarray*}
We now compute the dual function $\Qdual_{\treedistvec,
\eparambar}(\tausymb) \defn \inf_{\tepvec \in \neweflatman}
\Lag_{\treedistvec, \eparambar}(\tepvec, \tausymb)$; this minimization
problem can be decomposed into separate problems on each tree as
follows:
\begin{equation}
\label{EqnInf}
\sum_{\treegr}
\treedist(\treegr) \; \inf_{\myproj{\eparam}{\treegr} \in
\neweflatman(\treegr)} \big [ \Partinf(\eparam(\treegr)) -
\inprod{\eparam(\treegr)}{\tausymb} \big] \; + \;
\inprod{\tausymb}{\eparambar}.
\end{equation}
The following lemma is key to computing this infimum:
\begin{lemma}
%
\label{LemIndicate}
The function
\begin{eqnarray}
\label{EqnIndicate}
f(\tausym) & = & \sup_{\eparam(\treegr) \in \neweflatman(\treegr)}
\big \{ \inprod{ \myproj{\eparam}{\treegr}}{\tausym} -
\Partinf(\myproj{\eparam}{\treegr}) \big \} 
\end{eqnarray}
has the explicit form
\begin{equation}
\begin{cases} 0 & \operatorname{if} \;  \;  \tausym 
\in \Locset(\graph; \treegr) \\
+ \infty & \operatorname{otherwise},
\end{cases} \qquad
\end{equation}
where 
\begin{multline*}
\Locset(\graph; \treegr) \defn \Big \{ \; \tausymb \in \real^\df_+ \;
\big | \; \sum_{j \in \statesp_s} \tausym_{s;j} = 1 \; \forall s \in
\vertex, \\
\sum_{j \in \statesp_s} \tausym_{st;jk} = \tausym_{t;k} \; \forall \;
(s,t) \in \edge(\treegr) \; \Big \}
\end{multline*}
\end{lemma}
\begin{proof}
See Appendix~\ref{AppIndicate}.
\end{proof} \hfill  \\
Using Lemma~\ref{LemIndicate}, the value of the infimum~\eqref{EqnInf}
will be equal to $\inprod{\taupar}{\eparambar}$ if $\taupar \in
\Locset(\graph; \treegr)$ for all $\treegr \in \supp(\treedistvec)$,
and $-\infty$ otherwise. Since every edge in the graph belongs to at
least one tree in $\supp(\treedistvec)$, we have $\cap_{\treegr \in
\supp(\treedistvec)} \Locset(\graph; \treegr) \; \equiv \;
\Locset(\graph)$, so that the dual function takes the form:
\[
\Qdual_{\treedistvec, \eparambar}(\tausymb) \; = \; \begin{cases}
\inprod{\taupar}{\eparambar} & \mbox{if $\taupar \in \Locset(\graph)$}
\\ - \infty & \mbox{otherwise}.
\end{cases}
\]
Thus, the dual optimal value is $\max_{\taupar \in \Locset(\graph)}
\inprod{\taupar}{\eparambar}$; by strong
duality~\cite{Bertsekas_nonlin}, this optimum is equal to the optimal
primal value~\eqref{EqnBasicProb}.
\end{proof} \hfill  \\
The equivalence guaranteed by the duality relation in
Theorem~\ref{ThmLagDual} is useful, because the dual problem has much
simpler structure than the primal problem.  For a general distribution
$\treedistvec$, the primal problem~\eqref{EqnBasicProb} entails
minimizing a sum of functions over all spanning trees of the graph,
which can be a very large collection.  In contrast, the dual program
on the RHS of equation~\eqref{EqnTreeRelaxB} is simply a linear program
(LP) over $\Locset(\graph)$, which is a relatively simple polytope.
(In particular, it can be characterized by $\order(\statenum
\spnodenum + \statenum^2 |\edge|)$ constraints, where $\statenum \defn
\max_s |\statesp_s|$.)

This dual LP~\eqref{EqnTreeRelaxB} also has a very natural
interpretation.  In particular, the set $\Locset(\graph)$ is an outer
bound on the marginal polytope $\Margset(\graph)$, since any valid
marginal vector must satisfy all of the constraints defining
$\Locset(\graph)$.  Thus, the dual LP~\eqref{EqnTreeRelaxB} is simply a
relaxation of the original LP~\eqref{EqnMargRep}, obtained by
replacing the original constraint set $\Margset(\graph)$ by the set
$\Locset(\graph)$ formed by local (node and edgewise) constraints.
Note that for a graph with cycles, $\Locset(\graph)$ is a strict
superset of $\Margset(\graph)$.  (In particular,
Example~\ref{ExaPseudoFrac} to follow provides an explicit
construction of an element $\taupar \in \Locset(\graph) \bk
\Margset(\graph)$.)  For this reason, we call any $\tausymb \in
\Locset(\graph)$ a {\em pseudomarginal} vector.

An additional interesting connection is that this polytope
$\Locset(\graph)$ is equivalent to the constraint set involved in the
Bethe variational principle which, as shown by Yedidia et
al.~\cite{Yedidia02}, underlies the sum-product algorithm.  In
addition, it is possible to recover this LP relaxation as the
``zero-temperature'' limit of an optimization problem based on a
convexified Bethe approximation, as discussed in our previous
work~\cite{Wainwright05}.  For binary variables, the linear
program~\eqref{EqnTreeRelaxB} can be shown to be equivalent to a
relaxation that has been studied in previous
work~\cite[e.g.,]{Hammer84,Boros90}.  The derivation given here
illuminates the critical role of graphical structure in controlling
the tightness of such a relaxation.  In particular, an immediate
consequence of our development is the following:
\begin{corollary}
\label{CorTreeExact}
The relaxation~\eqref{EqnTreeRelaxB} is exact for any problem on a
tree-structured graph.
\end{corollary}

\begin{figure*}[t]

\bec
\begin{tabular}{ccccc}
\psfrag{#Tmargset#}{$\Locset(\graph)$}
\psfrag{#Margset#}{$\Margset(\graph)$}
\psfrag{#mufrac#}{$\meanpar_{frac}$}
\psfrag{#muint#}{$\meanpar_{int}$}
\widgraph{.28\textwidth}{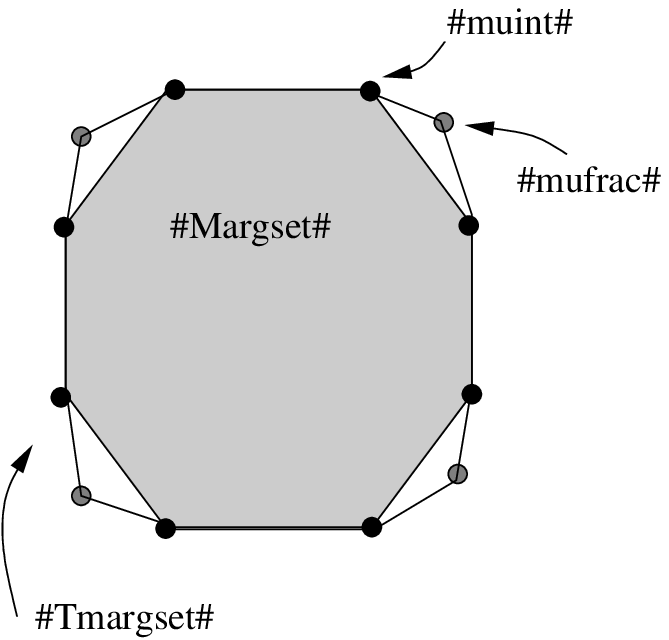} &
\hspace*{.1in} &
\psfrag{#eparam#}{$\eparambar^1$} \psfrag{#Margset#}{$\Margset(\graph)$}
\raisebox{.27in}{\widgraph{.22\textwidth}{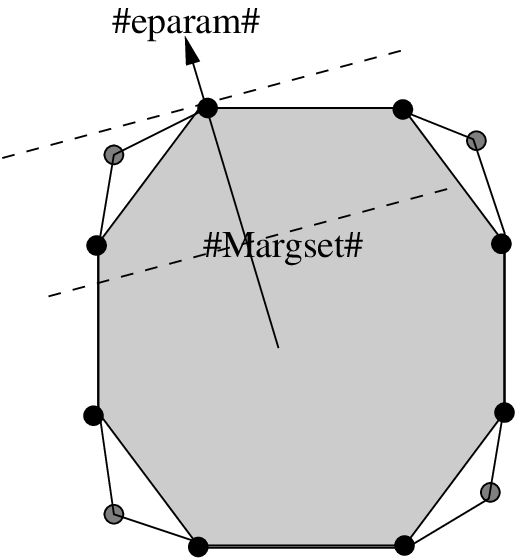}} &
\hspace*{.1in} &
\psfrag{#eparam#}{$\eparambar^2$}
\psfrag{#Margset#}{$\Margset(\graph)$}
\hspace*{.4in} \raisebox{.27in}{\widgraph{.29\textwidth}{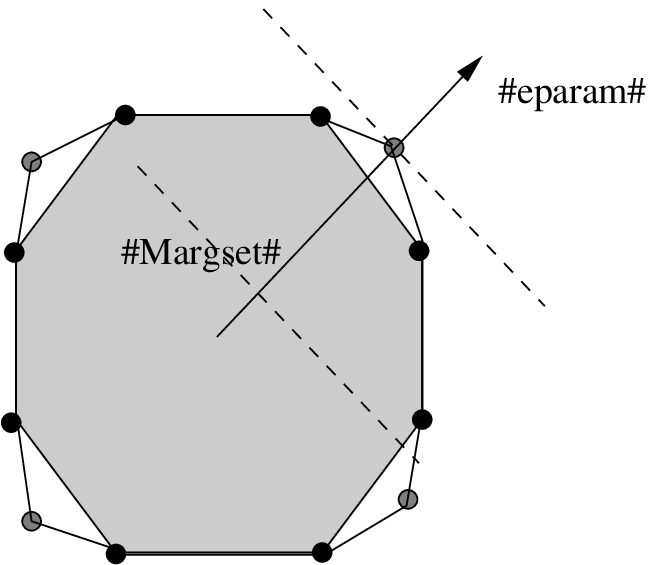}} \\
(a) & & (b) & & (c) \\
\end{tabular}
\caption{(a) The constraint set $\Locset(\graph)$ is an outer bound on
the exact marginal polytope.  Its vertex set includes all the integral
vertices of $\Margset(\graph)$, which are in one-to-one correspondence
with optimal solutions of the integer program.  It also includes
additional fractional vertices, which are {\em not} vertices of
$\Margset(\graph)$.  (b)-- (c) Solving a LP with cost vector
$\eparambar$ entails translating a hyperplane with normal $\eparambar$
until it is tangent to the constraint set $\Locset(\graph)$.  In (b),
the point of tangency occurs at a unique integral vertex.  In (c), the
tangency occurs at a fractional vertex of $\Locset(\graph)$ that lies
outside of $\Margset(\graph)$.}
\label{FigLocsetFrac}
\enc
\end{figure*}

Since the LP relaxation~\eqref{EqnTreeRelaxB} is always exact for MAP
estimation with any tree-structured distribution, we refer to it as a
{\em tree relaxation}.  For a graph with cycles --- in sharp contrast
to the tree-structured case ---- $\Locset(\graph)$ is a strict outer
bound on $\Margset(\graph)$, and the relaxation~\eqref{EqnTreeRelaxB}
will not always be tight.  Figure~\ref{FigLocsetFrac} provides an
idealized illustration of $\Locset(\graph)$, and its relation to the
exact marginal polytope $\Margset(\graph)$.  It can be seen that the
vertices of $\Margset(\graph)$ are all of the form
$\vmean{\meanpar}{\jcon}$, corresponding to the marginal vector
realized by the delta distribution that puts all its mass on $\jcon
\in \stsp$.  In the canonical overcomplete
representation~\eqref{EqnOver}, each element of any such
$\vmean{\meanpar}{\jcon}$ is either zero or one.  These {\em integral}
vertices, denoted by $\verint$, are drawn with black circles in
Figure~\ref{FigLocsetFrac}(a).  It is straightforward to show that
each such $\vmean{\meanpar}{\jcon}$ is also a vertex of
$\Locset(\graph)$.  However, for graphs with cycles, $\Locset(\graph)$
includes additional {\em fractional} vertices that lie strictly
outside of $\Margset(\graph)$, and that are drawn in gray circles in
Figure~\ref{FigLocsetFrac}(a).

\vtiny Since $\Locset(\graph)$ is also a polytope, the optimum of the
LP relaxation~\eqref{EqnTreeRelaxB} will be attained at a vertex
(possibly more than one) of $\Locset(\graph)$.  Consequently, solving
the LP relaxation using $\Locset(\graph)$ as an outer bound on
$\Margset(\graph)$ can have one of two possible outcomes.  The first
possibility is that optimum is attained at some vertex of
$\Locset(\graph)$ that is \emph{also} a vertex of $\Margset(\graph)$.
The optimum may occur at a unique integral vertex, as illustrated in
panel (b), or at multiple integral vertices (not illustrated here).
In this case, both the dual LP relaxation~\eqref{EqnTreeRelaxB}, and
hence also the primal version in equation~\eqref{EqnBasicUpper}, are
tight, and we can recover an optimal MAP configuration for the
original problem, which is consistent with
Proposition~\ref{PropTightBound}.  Alternatively, the optimum is
attained only outside the original marginal polytope
$\Margset(\graph)$ at a fractional vertex of $\Locset(\graph)$, as
illustrated in panel (c).  In this case, the relaxation must be loose,
so that Proposition~\ref{PropTightBound} asserts that it is impossible
to find a configuration that is optimal for all tree-structured
problems. Consequently, whether or not the tree agreement condition of
Proposition~\ref{PropTightBound} can be satisfied corresponds
precisely to the distinction between integral and fractional vertices
in the LP relaxation~\eqref{EqnTreeRelaxB}.

\begin{example}[Integral versus fractional vertices]
\label{ExaPseudoFrac}
In order to demonstrate explicitly the distinction between fractional
and integral vertices, we now consider the simplest possible example
--- namely, a binary problem $\xvec \in \{0,1\}^3$ defined on the
3-node cycle.  Consider the parameter vector $\eparambar$ with
components defined as follows:
\begin{subequations}
\begin{eqnarray}
\label{EqnDefnEpfail}
\eparambar_s & \defn & \begin{bmatrix} \eparambar_{s;0} &
\eparambar_{s;1}
\end{bmatrix} \; \defn \; \begin{bmatrix} 0 & 0 \end{bmatrix} \;
\forall \; s  \\
 \eparambar_{st} & = &
\begin{bmatrix} \eparambar_{st;00} & \eparambar_{st;01} \\
\eparambar_{st;10} & \eparambar_{st;11}
\end{bmatrix}
\; \defn \; \begin{bmatrix} 0 & -\cpar \\ -\cpar &
0
\end{bmatrix} \quad \forall \; (s,t).
\end{eqnarray}
\end{subequations}
Suppose first that $\cpar$ is positive --- say $\cpar = 1$.  By
construction of $\eparambar$, we have $\inprod{\eparambar}{\taupar}
\leq 0$ for all $\taupar \in \Locset(\graph)$.  This inequality is
tight when $\taupar$ is either the vertex $\vmean{\meanpar}{0}$
corresponding to the configuration $[0 \; 0 \; 0]$, or its counterpart
$\vmean{\meanpar}{1}$ corresponding to $[1 \; 1 \; 1 ]$.  In fact,
both of these configurations are MAP-optimal for the original problem,
so that we conclude that the LP relaxation~\eqref{EqnTreeRelaxB} is
tight (i.e., we can achieve tree agreement).

On the other hand, suppose that $\cpar < 0$; for concreteness, say
$\cpar = -1$.  This choice of $\eparambar$ encourages all pairs of
configurations $(x_s, x_t)$ to be distinct (i.e., $x_s \neq x_t$).
However, in going around the cycle, there must hold $x_s = x_t$ for at
least one pair.  Therefore, the set of optimal configurations consists
of $[1 \; 0 \; 1]$, and the other five permutations thereof.  (I.e.,
all configurations except $[1 \; 1 \; 1]$ and $[0 \; 0 \; 0]$ are
optimal). The value of any such optimizing configuration --- i.e.,
$\max_{\meanpar \in \Margset(\graph)} \inprod{\eparambar}{\meanpar}$
--- will be $-2 \cpar > 0$, corresponding to the fact that two of the
three pairs are distinct.

However, with reference to the relaxed polytope $\Locset(\graph)$, a
larger value of $\inprod{\eparambar}{\taupar}$ can be attained.  We
begin by observing that $\inprod{\eparambar}{\taupar} \leq -3 \cpar$
for all $\taupar \in \Locset(\graph)$.  In fact, equality can be
achieved in this inequality by the following pseudomarginal:
\begin{subequations}
\begin{eqnarray}
\label{EqnPseudoFrac}
\taupar_s & = & \begin{bmatrix} \taupar_{s;0} & \taupar_{s;1}
\end{bmatrix} \; \defn \; \begin{bmatrix} 0.5 & 0.5 \end{bmatrix} \;
\forall \; s,  \\
\taupar_{st} & = &
\begin{bmatrix} \taupar_{st;00} & \taupar_{st;01} \\
\taupar_{st;10} & \taupar_{st;11}
\end{bmatrix}
\; \defn \; \begin{bmatrix} 0 & 0.5 \\ 0.5 &
0
\end{bmatrix} \quad \forall \; (s,t).
\end{eqnarray}
\end{subequations}
Overall, we have shown that $\max_{\taupar \in \Locset(\graph)}
\inprod{\eparambar}{\taupar} \; = \; - 3\cpar \; > \; -2 \cpar \; = \;
\max_{\meanpar \in \Margset(\graph)} \inprod{\eparambar}{\meanpar}$,
which establishes that the relaxation~\eqref{EqnTreeRelaxB} is loose
for this particular problem.  Moreover, the pseudomarginal vector
$\tausym$ defined in equation~\eqref{EqnPseudoFrac} corresponds to a
fractional vertex of $\Locset(\graph)$, so that we are in the
geometric setting of Figure~\ref{FigLocsetFrac}(c). \hfill \finex
\end{example}

\section{Tree-reweighted message-passing algorithms}
\label{SecTrwMax}

The main result of the preceding section is that the problem of
finding tight upper bounds, as formulated in
equation~\eqref{EqnBasicProb}, is equivalent to solving the relaxed
linear program~\eqref{EqnTreeRelaxB} over the constraint set
$\Locset(\graph)$.  A key property of this constraint set is that it
is defined by a number of constraints that is at most quadratic in the
number of nodes $\spnodenum$.  Solving an LP over $\Locset(\graph)$,
then, is certainly feasible by various generic methods, including the
simplex algorithm~\cite[e.g.,]{Bertsimas}.  It is also of interest to
develop algorithms that exploit the graphical structure intrinsic to
the problem.  Accordingly, this section is devoted to the development
of iterative algorithms with this property.  An interesting property
of the iterative methods developed here is that when applied to a
tree-structured graph, they all reduce to the ordinary max-product
algorithm~\cite{FreemanMAP01,Wainwright02b}.  For graphs with cycles,
in contrast, they remain closely related to but nonetheless differ
from the ordinary max-product algorithm in subtle but important ways.
Ultimately, we establish a connection between particular fixed points
of these iterative algorithms and optimal dual solutions of the LP
relaxation~\eqref{EqnTreeRelaxB}.  In this way, we show that (suitably
reweighted) forms of the max-product algorithm have a variational
interpretation in terms of the LP relaxation.  As a corollary, our
results show that the ordinary max-product algorithm for trees (i.e.,
the Viterbi algorithm) can be viewed as an iterative method for
solving a particular linear program.

We begin with some background on the notion of {\em max-marginals},
and their utility in computing exact MAP estimates of tree-structured
distributions~\cite{Aji97,Cowell,Dawid92,Wainwright02b}.  We then
define an analogous notion of {\em pseudo-max-marginals} for graphs
with cycles, which play a central role in the message-passing
algorithms that we develop subsequently.

\subsection{Max-marginals for tree-distributions}

Although the notion of max-marginal can be defined for any
distribution, of particular interest in the current context are the
max-marginals associated with a distribution $p(\xvec;
\eparam(\treegr))$ that is Markov with respect to some tree $\treegr =
(\vertex, \edge(\treegr))$.  For each $s \in \vertex$ and $j \in
\statesp_s$, the associated single node max-marginal is defined by a
maximization over all other nodes in the graph
\begin{eqnarray}
\label{EqnDefnSingMaxmarg}
\grmaxmarg_{s;j} & \defn & \alphnorm_s \; \max_{ \{ \xvec \; | \; x_s =
j\} } p(\xvec; \eparam(\treegr)).
\end{eqnarray}
Here $\alphnorm_s > 0$ is some normalization constant, included for
convenience, that is independent of $j$ but can vary from node to
node.  Consequently, the max-marginal $\grmaxmarg_{s;j}$ is
proportional to the probability of the most likely configuration under
the constraint $x_s = j$. Note that $\grmaxmarg_{s;j}$ is obtained by
maximizing over the random variables at all nodes $t \neq s$, whence
the terminology ``max-marginal''.  For each edge $(s,t)$, the joint
pairwise max-marginal is defined in an analogous manner:
\begin{eqnarray}
\label{EqnDefnJointMaxmarg}
\grmaxmarg_{st;jk} & \defn & \alphnorm_{st} \; \max_{ \{ \xvec \; | \;
(x_s, x_t) = (j,k) \} } p(\xvec; \eparam(\treegr)).
\end{eqnarray}
Once again, the quantity $\alphnorm_{st}$ is a positive normalization
constant that can vary from edge to edge but does not depend on
$(j,k)$.

  It is convenient to represent all the values
$\{\grmaxmarg_{s;j}, j
\in \statesp_s\}$ associated with a given node, and the values
$\{\grmaxmarg_{st;jk}, (j,k) \in \statesp_s \times \statesp_t \}$
associated with a given edge in the functional form:
\begin{subequations}
\begin{eqnarray*}
\grmaxmarg_s(x_s) &  \defn & \sum_{j \in \statesp_s} \grmaxmarg_{s;j}
\Myind_j(x_s),  \\
\grmaxmarg_{st}(x_s, x_t) & \defn & \sum_{(j,k) \in \statesp_s \times
\statesp_t} \grmaxmarg_{st;jk} \Myind_j(x_s) \Myind_k(x_t).
\end{eqnarray*}
\end{subequations}
It is well-known~\cite{Cowell} that any tree-structured distribution
 $p(\xvec; \eparam(\treegr))$ can be factorized in terms of its
 max-marginals as follows:
\begin{eqnarray}
\label{EqnMaxMargRep}
p(\xvec; \eparam(\treegr)) & \propto & \prod_{s \in \vertex}
\grmaxmarg_s(x_s) \; \; \prod_{(s,t) \in \edge(\treegr)}
\frac{\grmaxmarg_{st}(x_s, x_t)}{\grmaxmarg_s(x_s) \grmaxmarg_t(x_t)}.
\end{eqnarray}
This factorization, which is entirely analogous to the more familiar
one in terms of (sum)-marginals, is a special case of the more general
junction tree decomposition~\cite{Dawid92,Cowell}.  Moreover, it can
be shown~\cite{Dawid92,Wainwright02b} that the ordinary max-product
(min-sum) algorithm computes this max-marginal factorization.  The
fact that this factorization can be computed in a straightforward
manner for any tree is exploited in the algorithms that we develop in
the sequel.

The max-marginal factorization~\eqref{EqnMaxMargRep} yields a local
criterion for assessing the validity of tree max-marginals.  The
following lemma provides a precise statement:
\begin{lemma}
\label{LemEdgeConsist}
A collection $\{\grmaxmarg_s, \grmaxmarg_{st} \}$ are valid
max-marginals for a tree if and only if the edgewise consistency
condition
\begin{eqnarray}
\label{EqnEdgeConsist}
\grmaxmarg_s(x_s) & = & \alphnorm \; \max_{x'_t \in \statesp_t}
\grmaxmarg_{st}(x_s, x'_t)
\end{eqnarray}
holds\footnote{Here $\alphnorm$ is a positive constant that depends on
both the edge, and the variable over which the maximization takes
place.} for every edge $(s,t) \in \edge(\treegr)$.
\end{lemma}
\begin{proof}
Necessity of the edge consistency is clear.  The sufficiency can be
established by an inductive argument in which successive nodes are
stripped from the tree by local maximization;
see~\cite{Dawid92,Wainwright02b} for further details.
\end{proof} \hfill  \\

The max-marginal representation~\eqref{EqnMaxMargRep} allows the
global problem of MAP estimation to be solved by performing a set of
local maximization operations.  In particular, suppose that the
configuration $\mxvec$ belongs to $\xoptset(\eparam(\treegr))$,
meaning that is MAP-optimal for $p(\xvec; \eparam(\treegr))$.  For a
tree, such configurations are completely characterized by {\em local
optimality conditions} with respect to the max-marginals, as
summarized in the following:
\begin{lemma}[Local optimality]
\label{LemLocalOpt}
Let $\{ \grmaxmarg_s, \grmaxmarg_{st} \}$ be a valid set of
max-marginals for a tree-structured graph. Then a configuration
$\mxvec$ belongs to $\xoptset(\eparam(\treegr))$ if and only if the
following local optimality conditions hold:
\begin{subequations}
\label{EqnLocalOpt}
\begin{eqnarray}
\label{EqnLocalOpta}
\mx_s \in \arg \max_{x_s} \grmaxmarg_s(x_s) & \forall & s, \\
\label{EqnLocalOptb}
(\mx_s, \mx_t) \in \arg \max_{x_s,x_t} \grmaxmarg_{st}(x_s,x_t) &
\forall & (s,t)
\end{eqnarray}
\end{subequations}
\end{lemma}
\begin{proof}
The necessity of the conditions in equation~\eqref{EqnLocalOpt} is
clear.  To establish sufficiency, we follow a dynamic-programming
procedure.  Any tree can be rooted at a particular node $r \in
\vertex$, and all edges can be directed from parent to child ($s
\rightarrow t$).  To find a configuration $\mxvec \in
\xoptset(\eparam(\treegr))$, begin by choosing an element $\xs_r \in
\arg \max_{x_r}
\grmaxmarg_r(x_r)$.  Then proceed recursively down the tree, from
parent $s$ to child $t$, at each step choosing the child configuration
$\xs_t$ from $\arg
\max_{x_t} \grmaxmarg_{st}(\xs_s, x_t)$.  By construction, the
configuration $\mxvec$ so defined is MAP-optimal;
see~\cite{Dawid92,Wainwright02b} for further details.
\end{proof} \hfill  \\
A particularly simple condition under which the local optimality
conditions~\eqref{EqnLocalOpt} hold is when for each $s \in \vertex$,
the max-marginal $\grmaxmarg_s$ has a {\em unique optimum} $\xs_s$.
In this case, the MAP configuration $\mxvec$ is unique with elements
$\xs_s =
\arg \max_{x_s} \grmaxmarg_s(x_s)$ that are computed easily by
maximizing each single node max-marginal.  If this uniqueness
condition does not hold, then more than one configuration is
MAP-optimal for $p(\xvec; \eparambar)$.  In this case, maximizing each
single node max-marginal is no longer sufficient~\cite{Wainwright02b},
and the dynamic-programming procedure described in the proof of
Lemma~\ref{LemLocalOpt} must be used.

\subsection{Iterative algorithms}

We now turn to the development of iterative algorithms for a graph
$\graph = (\vertex, \edge)$ that contains cycles.  We begin with a
high-level overview of the concepts and objectives, before proceeding
to a precise description.

\subsubsection{High level view}

The notion of max-marginal is not limited to distributions defined by
tree-structured graphs, but can also be defined for graphs with
cycles.  Indeed, if we were able to compute the exact max-marginals of
$p(\xvec; \eparambar)$ and each single node max-marginal had a unique
optimum, then the MAP estimation problem could be solved by local
optimizations.\footnote{If a subset of the single node max-marginals
had multiple optima, the situation would be more complicated.}
However, computing exact max-marginals for a distribution on a general
graph with cycles is an intractable task.  Therefore, it is again
necessary to relax our requirements. 

The basic idea, then, is to consider a vector of so-called {\em
pseudo-max-marginals} \mbox{$\grmaxmarg \defn \{
\grmaxmarg_s, \grmaxmarg_{st} \}$,} the properties of which are to be
defined shortly.  The qualifier ``pseudo'' reflects the fact that
these quantities no longer have an interpretation as exact
max-marginals, but instead represent approximations to max-marginals
on the graph with cycles.  For a given distribution $\treedistvec$
over the spanning trees of the graph $\graph$ and a tree $\treegr$ for
which $\treedist(\treegr) > 0$, consider the subset
$\meontree{\grmaxmarg}{\treegr}$, corresponding to those elements of
$\grmaxmarg$ associated with $\treegr$ --- i.e.,
\begin{eqnarray}
\label{EqnDefnMeontree}
\meontree{\grmaxmarg}{\treegr} & \defn & \{ \grmaxmarg_s, \; s \in
\vertex \} \; \cup \; \{\grmaxmarg_{st}, \; (s,t) \in \edge(\treegr)
\}.
\end{eqnarray}

We think of $\meontree{\grmaxmarg}{\treegr}$ as implicitly specifying
a tree-structured exponential parameter via the
factorization~\eqref{EqnMaxMargRep}, i.e.:
\begin{eqnarray}
\label{EqnTreeDet}
\meontree{\grmaxmarg}{\treegr} & \longleftrightarrow &
\meontree{\eparam}{\treegr}, \quad \forall \; \treegr \in \supp(\treedistvec),
\end{eqnarray}
which in turn implies that $\grmaxmarg$ is associated with a
collection of tree-structured parameters --- viz.:
\begin{eqnarray}
\label{EqnFullDet}
\grmaxmarg & \longleftrightarrow & \tepvec \defn \{
\meontree{\eparam}{\treegr} \; | \; \treegr \in \supp(\treedistvec) \}.
\end{eqnarray}
Now suppose that given $\treedistvec$, we have a vector $\grmaxmarg$
that satisfies the following properties:
\ben
\item[(a)]  The vector
$\grmaxmarg$ specifies a vector $\tepvec \in
\feaset_\treedistvec(\eparambar)$, meaning that $\tepvec$ is a {\em
$\treedistvec$-reparameterization} of the original distribution.
\item[(b)] For all trees $\treegr \in \supp(\treedistvec)$, the vector
$\meontree{\grmaxmarg}{\treegr}$ consists of exact max-marginals for
$p(\xvec;
\eparam(\treegr))$; we refer to this condition as {\em tree consistency}.
\een 
Our goal is to iteratively adjust the elements of $\grmaxmarg$ --- and
hence, implicitly, $\tepvec$ as well --- such that the
$\treedistvec$-reparameterization condition {\em always} holds, and
the tree consistency condition is achieved upon convergence.  In
particular, we provide algorithms such that any fixed point
$\grmaxmarg^*$ satisfies both conditions (a) and (b).

The ultimate goal is to use $\grmaxmarg^*$ to obtain a MAP
configuration for the target distribution $p(\xvec; \eparambar)$.  The
following condition turns out to be critical in determining whether or
not $\grmaxmarg^*$ is useful for this purpose: \\
\noindent {\bf {Optimum specification:}} The pseudo-max-marginals
$\{\grmaxmarg_s^*, \grmaxmarg_{st}^* \}$ satisfy the optimum
specification (OS) criterion if there exists at least one
configuration $\mxvec$ that satisfies the local optimality
conditions~\eqref{EqnLocalOpt} for every vertex $s \in \vertex$ and
edge $(s,t) \in \edge$ on the graph with cycles.

\noindent Note that the OS criterion always holds for any set of {\em
exact} max-marginals on any graph.  For the {\em pseudo}-max-marginals
updated by the message-passing algorithms, in contrast, the OS
criterion is no longer guaranteed to hold, as we illustrate in
Example~\ref{ExaDefnOSfail} to follow.

In the sequel, we establish that when $\grmaxmarg^*$ satisfies the OS
criterion with respect to some configuration $\mxvec$, then any such
$\mxvec$ must be MAP-optimal for the target distribution.  In
contrast, when the OS criterion is not satisfied, the
pseudo-max-marginals $\{\grmaxmarg^*_s, \grmaxmarg^*_{st}\}$ do not
specify a MAP-optimal configuration, as can be seen by a continuation
of Example~\ref{ExaPseudoFrac}.
\begin{example}[Failure of the OS criterion]
\label{ExaDefnOSfail}
Consider the parameter vector $\eparambar$ defined in
equation~\eqref{EqnDefnEpfail}.  Let the spanning tree distribution
$\treedistvec$ place mass $1/3$ on each of the three spanning trees
associated with a 3-node single cycle.  With this choice, the edge
appearances probabilities are $\treedist_{st} = 2/3$ for each edge.
We now define a 2-vector $\log \grmaxmarg^*_s$ of log
pseudo-max-marginals associated with node $s$, as well as a $2 \times
2$ matrix $\log \grmaxmarg^*_{st}$ of log pseudo-max-marginals
associated with edge $(s,t)$, in the following way:
\begin{subequations}
\begin{eqnarray*}
\log \grmaxmarg^*_s & \defn & \begin{bmatrix} 0 & 0 \end{bmatrix}
\quad \forall \; s \in \vertex \\
\log \grmaxmarg^*_{st} & \defn & \frac{1}{(2/3)} \;
\begin{bmatrix} 0 & -\cpar \\ -\cpar & 0 \end{bmatrix} \quad \forall
\; (s,t) \in \edge.
\end{eqnarray*}
\end{subequations}
For each of the three trees in $\supp(\treedistvec)$, the associated
vector $\meontree{\grmaxmarg^*}{\treegr}$ of pseudo-max-marginals
defines a tree-structured exponential parameter
$\meontree{\eparams}{\treegr}$ as in equation~\eqref{EqnTreeDet}.
More specifically, we have
\begin{subequations}
\begin{eqnarray*}
\meontree{\eparam^*_s}{\treegr} & = &  \log \grmaxmarg^*_s \quad
\forall \; s\ \in \vertex, \\
\meontree{\eparam^*_{st}}{\treegr} & = & \begin{cases} \log
\grmaxmarg^*_{st} & \forall \; (s,t) \in \edge(\treegr) ,\\ 0 &
\mbox{otherwise}
\end{cases}
\end{eqnarray*}
\end{subequations}
With this definition, it is straightforward to verify that
$\sum_{\treegr} \frac{1}{3} \eparams(\treegr) = \eparambar$, meaning
that the $\treedistvec$-reparameterization condition holds.  Moreover,
for any $\cpar \in \real$, the edgewise consistency condition $\max_{x'_t}
\grmaxmarg^*_{st}(x_s, x_t) = \alphnorm \;
\grmaxmarg^*_s(x_s)$ holds.  Therefore, the pseudo-max-marginals are
pairwise-consistent, so that by Lemma~\ref{LemEdgeConsist}, they are
tree-consistent for all three spanning trees.

Now suppose that $\cpar > 0$.  In this case, the pseudo-max-marginal
vector $\grmaxmarg^*$ does satisfy the OS criterion.  Indeed, both
configurations $[0 \; 0 \; 0]$ and $[1 \; 1 \; 1]$ achieve
$\argmax_{x_s} \grmaxmarg^*_s(x_s)$ for all vertices $s \in \vertex$,
and $\argmax_{x_s, x_t} \grmaxmarg^*_{st}(x_s, x_t)$ for all edges
$(s,t) \in \edge$.  This finding is consistent with
Example~\ref{ExaPseudoFrac}, where we demonstrated that both
configurations are MAP-optimal for the original problem, and that the
LP relaxation~\eqref{EqnTreeRelaxB} is tight.

Conversely, suppose that $\cpar < 0$.  In this case, the requirement
that $\mxvec$ belong to the set $\argmax_{x_s, x_t}
\grmaxmarg^*_{st}(x_s, x_t)$ for {\em all} three edges means that
$\xs_s \neq \xs_t$ for all three pairs.  Since this condition cannot
be met, the pseudo-max-marginal fails the OS criterion for $\cpar <
0$.  Again, this is consistent with Example~\ref{ExaPseudoFrac}, where
we found that for $\cpar < 0$, the optimum of the LP
relaxation~\eqref{EqnTreeRelaxB} was attained only at a fractional
vertex.  \hfill \finex
\end{example}

\subsubsection{Direct updating of pseudo-max-marginals}

Our first algorithm is based on updating a collection of
pseudo-max-marginals $\{\grmaxmarg_s, \grmaxmarg_{st} \}$ for a graph
with cycles such that $\treedistvec$-reparameterization (condition
(a)) holds at every iteration, and tree consistency (condition (b)) is
satisfied upon convergence. At each iteration $n=0,1,2, \ldots$,
associated with each node $s \in \vertex$ is a single node
pseudo-max-marginal $\grmaxmarg^n_s$, and with each edge $(s,t) \in
\edge$ is a joint pairwise
pseudo-max-marginal $\grmaxmarg^n_{st}$.  Suppose that for each tree
$\treegr$ in the support of $\treedistvec$, we use these
pseudo-max-marginals $\{\grmaxmarg^n_s, \grmaxmarg^n_{st} \}$ to
define a tree-structured exponential parameter $\eparam^n(\treegr)$
via equation~\eqref{EqnMaxMargRep}.  More precisely, again using the
functional notation as in equation~\eqref{EqnFuncNote}, the
tree-structured parameter $\eparam^n(\treegr)$ is defined in terms of
(element-wise) logarithms of $\grmaxmarg^n$ as follows:
\begin{subequations}
\label{EqnPseudo2Eparam}
\begin{eqnarray}
\label{EqnPseudo2Eparama}
\eparam^n_s(\treegr)(x_s) & = & \log \grmaxmarg^n_s(x_s) \qquad
\forall \; \; s \in \vertex \\
\label{EqnPseudo2Eparamb}
\eparam^n_{st}(\treegr)(x_s, x_t) & = & \begin{cases} \log
\frac{\grmaxmarg_{st}^n(x_s, x_t)}{\grmaxmarg^n_s(x_s)
\grmaxmarg^n_t(x_t)} & \mbox{if $(s,t) \in \edge(\treegr)$} \\
0 & \mbox{otherwise}
\end{cases}
\end{eqnarray}
\end{subequations}

The general idea is to update the pseudo-max-marginals iteratively, in
such a way that the $\treedistvec$-reparameterization condition is
maintained, and the tree consistency condition is satisfied upon
convergence. There are a number of ways in which such updates can be
structured; here we distinguish two broad classes of strategies:
tree-based updates, and parallel edge-based updates.  {\em Tree-based
updates} entail performing multiple iterations of updating on a fixed
tree $\treegr \in \supp(\treedistvec)$, updating only the 
subcollection $\meontree{\grmaxmarg}{\treegr}$ of pseudo-max-marginals
associated with vertices and edges in $\treegr$ until it is fully
tree-consistent {\em for this} tree (i.e., so that the components of
$\meontree{\grmaxmarg}{\treegr}$ are indeed max-marginals for the
distribution $p(\xvec; \meontree{\eparam}{\treegr})$).  However, by
focusing on this one tree, we may be changing some of the
$\grmaxmarg_s$ and $\grmaxmarg_{st}$ so that we do {\em not} have
tree-consistency on one or more of the other trees $\treegr' \in
\supp(\treedistvec)$.  Thus, the next step entails updating the
pseudo-max-marginals $\meontree{\grmaxmarg}{\treegr'}$ on one of the
other trees and so on, until ultimately the full collection
$\grmaxmarg$ is consistent on every tree.  In contrast, the {\em
edge-based strategy} involves updating the pseudo-max-marginal
$\grmaxmarg_{st}$ on each edge, as well as the associated single node
max-marginals $\grmaxmarg_s$ and $\grmaxmarg_t$, in parallel.  This
edgewise strategy is motivated by Lemma~\ref{LemEdgeConsist}, which
guarantees that $\grmaxmarg$ is consistent on every tree of the graph
if and only if the edge consistency condition
\begin{eqnarray}
\label{EqnEdge2}
\max_{x'_t \in \statesp_t} \grmaxmarg_{st}(x_s, x'_t) & = & \alphnorm
\; \grmaxmarg_s(x_s)
\end{eqnarray}
holds for every edge $(s,t)$ of the graph with cycles.

It should be noted that tree-based updates are computationally
feasible only when the support of the spanning tree distribution
$\treedistvec$ consists of a manageable number of trees.  When
applicable, however, there can be important practical benefits to
tree-based updates, including more rapid convergence as well as the
possibility of determining a MAP-optimal configuration {\em prior} to
convergence.  More details on tree-based updates and their properties
in more detail in Appendix~\ref{AppTreeBase}.  We provide some
experimental results demonstrating the advantages of tree-based
updates in Section~\ref{SecAdd}.

Here we focus on edge-based updates, due their simplicity and close
link to the ordinary max-product algorithm that will be explored in
the following section.  The edge-based reparameterization algorithm
takes the form shown in Figure~\ref{AlgEdgeRepar}; a few properties
are worthy of comment.  First, each scalar $\treedist_{st}$ appearing
in equations~\eqref{EqnInitReparamb} and~\eqref{EqnDefnMultUpa} is the
edge appearance probability of edge $(s,t)$ induced by the spanning
tree distribution $\treedistvec$, as defined in
Section~\ref{SecConvexTree}.  Second, this edge-based
reparameterization algorithm is very closely related to the ordinary
max-product algorithm~\cite{Wainwright02b}.  In fact, if
$\treedist_{st} = 1$ for all edges $(s,t) \in \edge$, then the
updates~\eqref{EqnDefnMultUp} are exactly equivalent to a
(reparameterization form) of the usual max-product updates.  We will
see this equivalence explicitly in our subsequent presentation of
tree-reweighted message-passing updates.

\bec
\begin{figure*}[t]
\framebox[0.98\textwidth]{\parbox{.95\textwidth}{
%
\noindent {\bf{Algorithm~\algrep: Edge-based reparameterization updates:}}
\hfill \\
\vspace*{.01in}

\noindent 1. Initialize the pseudo-max-marginals $\{ \grmaxmarg^0_s,
\grmaxmarg^0_{st} \}$ in terms of the original exponential parameter
vector as follows:
\begin{subequations}
\label{EqnInitReparam}
\begin{eqnarray}
\label{EqnInitReparama}
\grmaxmarg^0_s(x_s) & = & \alphnorm \; \exp \big(\eparambar_s(x_s) \big) \\
\label{EqnInitReparamb}
\grmaxmarg^0_{st}(x_s, x_t) & = & \alphnorm \; \exp
\big(\frac{1}{\treedist_{st}} \eparambar_{st}(x_s, x_t) +
\eparambar_t(x_t) + \eparambar_s(x_s) \big)
\end{eqnarray}
\end{subequations}
\vspace*{.01in} \\
\noindent 2. For iterations $n = 0, 1, 2, \ldots$, update the
pseudo-max-marginals as follows:
\begin{subequations}
\label{EqnDefnMultUp}
\begin{eqnarray}
\label{EqnDefnMultUpa}
\grmaxmarg^{n+1}_s(x_s) & = & \alphnorm \; \grmaxmarg^n_s(x_s) \;
\prod_{t \in \neigh(s)} \Big [ \frac{\max_{x'_t} \grmaxmarg^n_{st}(x_s,
x'_t)}{\grmaxmarg^n_s(x_s)} \Big]^{\treedist_{st}} \\
\label{EqnDefnMultUpb}
\grmaxmarg^{n+1}_{st}(x_s, x_t) & = & \alphnorm \;
\frac{\grmaxmarg^n_{st}(x_s,x_t)}{\max_{x'_t} \grmaxmarg^n_{st}(x_s,
x'_t) \; \; \max_{x'_s} \grmaxmarg^n_{st}(x'_s, x_t)} \; \;
\grmaxmarg^{n+1}_s(x_s) \grmaxmarg_t^{n+1}(x_t)
\end{eqnarray}
\end{subequations}
%
%
}}
\caption{Edge-based reparameterization updates of the
pseudo-max-marginals.}
\label{AlgEdgeRepar}
\end{figure*}
\enc

The following lemmas summarize the key properties of the
Algorithm~\ref{AlgEdgeRepar}.  We begin by claiming that all iterates
of this algorithm specify a $\treedistvec$-reparameterization:
\begin{lemma}[$\treedistvec$-reparameterization]
\label{LemReparamRep}
At each iteration $n = 0,1,2, \ldots$, the collection of
tree-structured parameter vectors $\tepvec^n = \{
\meontree{\eparam^n}{\treegr} \}$, as specified by the
pseudo-max-marginals $\{ \grmaxmarg_s^n, \grmaxmarg^n_{st} \}$ via
equation~\eqref{EqnPseudo2Eparam}, satisfies the
$\treedistvec$-reparameterization condition.
\end{lemma}
\begin{proof}
Using the initialization~\eqref{EqnInitReparam} and
equation~\eqref{EqnPseudo2Eparam}, for each tree $\treegr \in
\supp(\treedistvec)$, we have the relation $\eparam^0_{st}(\treegr) =
\eparambar_{st}/\treedist_{st}$ for all edges $(s,t) \in
\edge(\treegr)$, and $\eparam^0_s(\treegr) = \eparambar_s$ for all
vertices $s \in \vertex$.  Thus we have $\sum_\treegr
\treedist(\treegr) \eparam^0_s(\treegr) = \eparambar_s$ for all $s \in
\vertex$ and $\sum_{\treegr} \treedist(\treegr)
\eparam^0_{st}(\treegr) = \sum_{\treegr \ni (s,t)} \treedist(\treegr)
\frac{\eparambar_{st}}{\treedist_{st}} = \eparambar_{st}$ for all
$(s,t) \in \edge$, so that $\treedistvec$-reparameterization holds for
$n=0$.  We now proceed inductively: supposing that it holds for
iteration $n$, we prove that it also holds for iteration $n+1$.  Using
the update equation~\eqref{EqnDefnMultUp} and
equation~\eqref{EqnPseudo2Eparam}, we find that for all $\xvec \in
\statesp$, the quantity $\eparam^{n+1}(\treegr)(\xvec)$ is equal to
\begin{multline*}
\sum_{s \in \vertex} \biggr \{ \log \grmaxmarg^n_s(x_s) + \sum_{t \in
\neigh(s)} \treedist_{st} \log \frac{\max_{x'_t}
\grmaxmarg^n_{st}(x_s, x'_t)}{\grmaxmarg^n_s(x_s)} \biggr \} + \\
\sum_{(s,t) \in \edge(\treegr)} \log
\frac{\grmaxmarg^n_{st}(x_s,x_t)}{\max \limits_{x'_t}
\grmaxmarg^n_{st}(x_s, x'_t) \; \; \max \limits_{x'_s}
\grmaxmarg^n_{st}(x'_s, x_t)} 
\end{multline*}
Some algebraic re-arrangement leads to an equivalent expression (up to
additive constants independent of $\xvec$) for the weighted sum
$\sum_{\treegr} \treedist(\treegr) \eparam^{n+1}(\treegr)(\xvec)$:
\[
\sum_{s \in \vertex} \log \grmaxmarg^n_s(x_s) + \sum_{(s,t) \in \edge}
\treedist_{st} \log \frac{\grmaxmarg^n_{st}(x_s, x_t)}{\grmaxmarg^n_s(x_s)
\grmaxmarg^n_t(x_t)},
\]
which, using equation~\eqref{EqnPseudo2Eparam}, is seen to be equal to
$\sum_{\treegr} \treedist(\treegr) \eparam^n(\treegr)(\xvec)$.  Thus,
the statement follows by the induction hypothesis.
\end{proof} \hfill  \\

\noindent Next we characterize the fixed points of the updates in step
2:
\begin{lemma}
Any fixed point $\grmaxmarg^*$ of the updates~\eqref{EqnDefnMultUp}
satisfies the tree consistency condition (b).
\end{lemma}
\begin{proof}
At a fixed point, we can substitute $\grmaxmarg^* = \grmaxmarg^n =
\grmaxmarg^{n+1}$ at all places in the updates.  Doing so in
equation~\eqref{EqnDefnMultUpb} and cancelling out common terms leads
to the relation
\[
\kappa \; \frac{\grmaxmarg^*_{st}(x_s)}{\max_{x'_t}
\grmaxmarg^*_{st}(x_s, x'_t)} \; \;
\frac{\grmaxmarg^*_t(x_t)}{\max_{x'_s} \grmaxmarg^*_{st}(x'_s, x_t)}
\; = \; 1
\]
for all $(x_s, x_t)$, from which the edgewise consistency
condition~\eqref{EqnEdge2} follows for each edge $(s,t) \in \edge$.
The tree consistency condition then follows from
Lemma~\ref{LemEdgeConsist}.
\end{proof} \hfill  \\

\subsubsection{Message-passing updates}

The reparameterization updates of Algorithm~\ref{AlgEdgeRepar} can
also be described in terms of explicit message-passing operations.  In
this formulation, the pseudo-max-marginals depend on the original
exponential parameter vector $\eparambar$, as well as a set of
auxiliary quantities $\mess_{st}(\cdot)$ associated with the edges of
$\graph$.  For each edge $(s,t) \in \edge$, $\mess_{st}: \statesp_t
\rightarrow \real_+$ is a function from the state space $\statesp_t$
to the positive reals.  The function $\mess_{st}(\cdot)$ represents
information that is relayed from node $s$ to node $t$, so that we
refer to it as a ``message''.  The resulting algorithm is an
alternative but equivalent implementation of the reparameterization
updates of Algorithm~\ref{AlgEdgeRepar}.

More explicitly, let us define pseudo-max-marginals $\{\grmaxmarg_s,
\grmaxmarg_{st} \}$ in terms of $\eparambar$ and a given set of
messages $\mess = \{ \mess_{st} \}$ as follows:
\begin{subequations}
\label{EqnDefnPmax}
\begin{eqnarray}
\label{EqnDefnPmaxa}
\grmaxmarg_s(x_s) & \propto & \exp \big ( \eparambar_s(x_s) \big ) \;
\prod_{v \in \neigh(s)} \big[\mess_{vs}(x_s) \big]^{\treedist_{vs}} \\
\label{EqnDefnPmaxb}
\grmaxmarg_{st}(x_s, x_t) & \propto & \varphi_{st}(x_s, x_t) \;
\frac{\prod \limits_{v \in \neigh(s) \bk t}
\big[\mess_{vs}(x_s)\big]^{\treedist_{vs}} }{ \big[ \mess_{ts}(x_s)
\big]^{(1-\treedist_{ts})}}  \nonumber \\
&  & \qquad \quad  \times \frac{\prod \limits_{v \in \neigh(t)
\bk s} \big[\mess_{vt}(x_t)\big]^{\treedist_{vt}} }{ \big[
\mess_{st}(x_t) \big]^{(1-\treedist_{st})}}.
\end{eqnarray}
\end{subequations}
where $\phi_{st}(x_s, x_t) \defn \exp \big(\frac{1}{\treedist_{st}}
\eparambar_{st}(x_s, x_t) + \eparambar_s(x_s) + \eparambar_t(x_t)
\big)$.  As before, these pseudo-max-marginals can be used to define a
collection of tree-structured exponential parameters $\tepvec =
\{\meontree{\eparam}{\treegr}\}$ via
equation~\eqref{EqnPseudo2Eparam}.  First, we claim that for any
choice of messages, the set of tree-structured parameters so defined
specifies a $\treedistvec$-reparameterization:
\begin{lemma}
\label{LemAdmissible}
For any choice of messages, the collection $\{\eparam(\treegr)\}$ is
a $\treedistvec$-reparameterization of $\eparambar$.
\end{lemma}
\begin{proof}
We use the definition~\eqref{EqnPseudo2Eparam} of
$\meontree{\eparam}{\treegr}$ in terms of $\{ \grmaxmarg_s,
\grmaxmarg_{st} \}$ to write $\sum_{\treegr} \treedist(\treegr)
\eparam(\treegr)(\xvec)$ as
\begin{equation*}
\sum_{\treegr}\treedist(\treegr) \Big [\sum_{s \in \vertex} \log
\grmaxmarg_s(x_s) + \sum_{(s,t) \in \edge(\treegr)} \log
\frac{\grmaxmarg_{st}(x_s, x_t)}{\grmaxmarg_s(x_s) \grmaxmarg_t(x_t) }
\Big].
\end{equation*}
Expanding out the expectation yields
\begin{equation}
\label{EqnTempSum}
\sum_{s \in \vertex} \log \grmaxmarg_s(x_s) + \sum_{(s,t) \in \edge}
\treedist_{st} \log \frac{\grmaxmarg_{st}(x_s, x_t)}{\grmaxmarg_s(x_s)
\grmaxmarg_t(x_t) },
\end{equation}
Using the definition~\eqref{EqnDefnPmax} of $\grmaxmarg_s$ and
$\grmaxmarg_{st}$, we have
\begin{multline*}
\treedist_{st} \log \frac{\grmaxmarg_{st}(x_s, x_t)}{\grmaxmarg_s(x_s)
\grmaxmarg_t(x_t)}  = \\\eparambar_{st}(x_s, x_t) - \treedist_{st}
\log \mess_{st}(x_t) - \treedist_{st} \log \mess_{ts}(x_s),
\end{multline*}
As a consequence, each weighted log message $\treedist_{st} \log
\mess_{ts}(x_s)$ appears twice in equation~\eqref{EqnTempSum}: once in
the term $\log \grmaxmarg_s(x_s)$ with a plus sign, and once in the
term $\log \grmaxmarg_{st}(x_s,x_t)/\grmaxmarg_s(x_s)
\grmaxmarg_t(x_t)$ with a negative sign.  Therefore, the messages all
cancel in the summation.  This establishes that for all $\xvec \in
\stsp$, we have $\sum_{\treegr} \treedist(\treegr)
\eparam(\treegr)(\xvec) = \sum_{s \in \vertex} \eparambar_s(x_s) +
\sum_{(s,t) \in \edge} \eparambar_{st}(x_s, x_t)$.
\end{proof} \hfill  \\
The message-passing updates shown in Figure~\ref{FigAlgTRW} are designed
to find a collection of pseudo-max-marginals $\{\grmaxmarg_s,
\grmaxmarg_{st} \}$ that satisfy the tree consistency condition (b).
\bec
\begin{figure*}[t]
\framebox[0.98\textwidth]{\parbox{.95\textwidth}{
{\bf{Algorithm~\algtrw: Parallel tree-reweighted max-product}} \hfill \\
\vspace*{.01in}

\noindent 1. Initialize the messages $\mess^0 = \{\mess^0_{st} \}$
with arbitrary positive real numbers. \\
\vspace*{.01in} \\
\noindent 2. For iterations $n = 0, 1, 2, \ldots$, update the messages
as follows: 
\begin{equation}
\label{EqnMessFix}
\mess^{n+1}_{ts}(x_s) = \alphnorm \; \max_{x'_t \in \statesp_t} \Big
\{ \exp \biggr (\frac{1}{\treedist_{st}} \eparambar_{st}(x_s, x'_t) +
\eparambar_t(x'_t) \biggr) \; \frac{\prod_{v \in \neigh(t) \bk s}
\big[\mess^n_{vt}(x'_t)\big]^{\treedist_{vt}} }{ \big[
\mess^n_{st}(x'_t) \big]^{(1-\treedist_{st})}} \Big \}
\end{equation}
}}
\caption{Parallel edge-based form of tree-reweighted message-passing
updates.  The algorithm reduces to the ordinary max-product updates
when all the edge weights $\treedist_{st}$ are set equal to one.}
\label{FigAlgTRW}
\end{figure*}
\enc

First, it is worthwhile noting that the message update
equation~\eqref{EqnMessFix} is closely related to the
standard~\cite{FreemanMAP01,Wainwright02b} max-product updates, which
correspond to taking $\treedist_{st} = 1$ for every edge.  On one
hand, if the graph $\graph$ is actually a tree, any vector in the
spanning tree polytope must necessarily satisfy $\treedist_{st} = 1$
for every edge $(s,t) \in \edge$, so that Algorithm~\algtrw$\:$ reduces to
the ordinary max-product update.  However, if $\graph$ has cycles,
then it is impossible to have $\treedist_{st} = 1$ for every edge
$(s,t) \in \edge$, so that the updates in equation~\eqref{EqnMessFix}
differ from the ordinary max-product updates in three critical ways.
To begin, the exponential parameters $\eparambar_{st}(x_s, x_t)$ are
scaled by the (inverse of the) edge appearance probability
$1/\treedist_{st} \; \geq \; 1$.  Second, for each neighbor $v \in
\neigh(t) \bk s$, the incoming message $\mess_{vt}$ is exponentiated
by the corresponding edge appearance probability $\treedist_{vt} \;
\leq \; 1$.  Last of all , the update of message $\mess_{ts}$ --- that
is, from $t$ to $s$ along edge $(s,t)$ --- depends on the {\em reverse
direction} message $\mess_{st}$ from $s$ to $t$ along the same
edge. Despite these features, the messages can still be updated in an
asynchronous manner, as in ordinary
max-product~\cite{FreemanMAP01,Wainwright02b}.

Moreover, we note that these the tree-reweighted updates are related
but distinct from the attenuated max-product updates proposed by Frey
and Koetter~\cite{Frey00}.  A feature common to both algorithms is the
re-weighting of messages; however, unlike the tree-reweighted
update~\eqref{EqnMessFix}, the attenuated max-product update
in~\cite{Frey00} of the message from $t$ to $s$ does not involve the
message in the reverse direction (i.e., from $s$ to $t$).

\vtiny

By construction, any fixed point of Algorithm~\algtrw$\:$ specifies a
set of tree-consistent pseudo-max-marginals, as summarized in the
following:
\begin{lemma}
\label{LemFixEdgeCon}
For any fixed point $\mess^*$ of the updates~\eqref{EqnMessFix}, the
associated pseudo-max-marginals $\grmaxmarg^*$ defined as in
equation~\eqref{EqnDefnPmaxa} and~\eqref{EqnDefnPmaxb} satisfy the
tree-consistency condition.
\end{lemma}
\begin{proof}
By Lemma~\ref{LemEdgeConsist}, it suffices to verify that the edge
consistency condition~\eqref{EqnEdge2} holds for all edges $(s,t) \in
\edge$.  Using the definition~\eqref{EqnDefnPmax} of $\grmaxmarg^*_s$
and $\grmaxmarg^*_{st}$, the edge consistency
condition~\eqref{EqnEdge2} is equivalent to equating $\exp \big (
\eparambar_s(x_s) \big ) \; \prod_{v \in \neigh(s)}
\big[\mess_{vs}(x_s) \big]^{\treedist_{vs}}$ with
\begin{multline*}
\alphnorm \; \max_{x'_t \in \statesp_t } \Big \{ \varphi_{st}(x_s,
x'_t) \; \frac{\prod_{v \in \neigh(s) \bk t}
\big[\mess_{vs}(x_s)\big]^{\treedist_{vs}} }{ \big[ \mess_{ts}(x_s)
\big]^{(1-\treedist_{ts})}} \\
\times \; \frac{\prod_{v \in \neigh(t) \bk s}
\big[\mess_{vt}(x'_t)\big]^{\treedist_{vt}} }{ \big[ \mess_{st}(x'_t)
\big]^{(1-\treedist_{st})}} \Big \}.
\end{multline*}
Pulling out all terms involving $\mess_{ts}(x_s)$, and canceling out
all remaining common terms yields the message update
equation~\eqref{EqnMessFix}.
\end{proof} \hfill  \\
\subsection{Existence and properties of fixed points}
\label{SecAnalFix}

We now consider various questions associated with Algorithms~\algrep$\:$
and~\algtrw, including existence of fixed points, convergence of the
updates, and the relation of fixed points to the LP
relaxation~\eqref{EqnTreeRelaxB}.  As noted previously, the two
algorithms (reparameterization and message-passing) represent
alternative implementations of the same updates, and hence are
equivalent in terms of their fixed point and convergence properties.
For the purposes of the analysis given here, we focus on the
message-passing updates given in Algorithm~\algtrw.

With reference to the first question, in related
work~\cite{Wainwright02b}, we proved the existence of fixed points for
the ordinary max-product algorithm when applied to any distribution
with strictly positive compatibilities defined on an arbitrary graph.
The same proof can be adapted to show that the message-update
equation~\eqref{EqnMessFix} has at least one fixed point $\mess^*$
under these same conditions.  Unfortunately, we do not yet have
sufficient conditions to guarantee convergence on graphs with cycles;
however, in practice, we find that the edge-based message-passing
updates~\eqref{EqnMessFix} converge if suitably damped. In particular,
we apply damping in the logarithmic domain, so that messages are
updated according to $\lambda \log \mess^{new}_{ts} + (1-\lambda) \log
\mess^{old}_{ts}$, where $\mess^{new}_{ts}$ is calculated in
equation~\eqref{EqnMessFix}.  Moreover, we note that in follow-up
work, Kolmogorov~\cite{Kol05} has developed a modified form of
tree-based updates for which certain convergence properties are
guaranteed.

Finally, the following theorem addresses the nature of the fixed
points, and in particular provides sufficient conditions for
Algorithm~\algtrw$\:$ to yield exact MAP estimates for the target
distribution $p(\xvec; \eparambar)$, and thereby establishes a link to
the dual of the LP relaxation of Theorem~\ref{ThmLagDual}:
\newcommand{\logmess}{\ensuremath{\omega}}
\newcommand{\newlag}{\ensuremath{\lambda}}

\begin{theorem}
\label{ThmExactMAP}
Let $\mess^*$ be a fixed point of Algorithm~\algtrw, and suppose
that the associated pseudo-max-marginals $\grmaxmargb^*$ satisfy the
optimum specification (OS) criterion. Then the following statements
hold:
\ben
\item[(a)] Any configuration $\xvec^*$ satisfying the local optimality
conditions in the OS criterion is a MAP configuration for $p(\xvec;
\eparambar)$.
\item[(b)] Let $\logmess^* = \log \mess^*$ be the logarithm of the
fixed point $\mess^*$ taken element-wise.  Then a linear combination
of $\logmess^*$ specifies an optimal solution to the dual of the LP
relaxation~\eqref{EqnTreeRelaxB} of Theorem~\ref{ThmLagDual}.
\een
\end{theorem}
\begin{proof}
\noindent (a) By Lemma~\ref{LemAdmissible}, the pseudo-max-marginals
$\grmaxmarg^*$ specify a $\treedistvec$-reparameterization of
$p(\xvec; \eparambar)$.  Since the message vector $\mess^*$ defining
$\grmaxmarg^*$ is a fixed point of the update
equation~\eqref{EqnMessFix}, Lemma~\ref{LemFixEdgeCon} guarantees that
the tree consistency condition holds.  By the optimum specification
(OS) criterion, we can find a configuration $\mxvec$ that is node and
edgewise optimal for $\grmaxmarg^*$.  By Lemma~\ref{LemLocalOpt}, the
configuration $\mxvec$ is optimal for every tree-structured
distribution $p(\xvec; \eparam^*(\treegr))$.  Thus, by
Proposition~\ref{PropTightBound}, the configuration $\mxvec$ is
MAP-optimal for $p(\xvec; \eparambar)$.  

\noindent (b) Let $\mess^*$ be a fixed point of the update
equation~\eqref{EqnMessFix}, such that the pseudo-max-marginals
$\grmaxmarg^*$ satisfy the OS criterion.  The proof involves showing
that a linear combination of the vector $\logmess^* \defn \log
\mess^*$, defined by the element-wise logarithm of the message fixed
point, is an optimal solution to a particular Lagrangian dual
reformulation of the LP relaxation~\eqref{EqnTreeRelaxB}.  For this
proof, it is convenient to represent any pseudomarginal $\taupar$ more
compactly in the functional form
\begin{subequations}
\begin{eqnarray*}
\taupar_s(x_s) & \defn & \sum_{j \in \statesp_s} \taupar_{s;j}
\Myind_j(x_s), \\
\taupar_{st}(x_s, x_t) & \defn & \sum_{(j,k) \in \statesp_s \times
\statesp_t} \taupar_{st;jk} \Myind_j(x_s) \Myind_k(x_t).
\end{eqnarray*}
\end{subequations}
For each edge $(s,t)$ and element $x_s \in \statesp_s$, define the
linear function $\contwo_{ts}(x_s) \defn \taupar_s(x_s) - \sum_{x'_t
\in \statesp_t} \taupar_{st}(x_s, x'_t)$, and let $\newlag_{ts}(x_s)$
be a Lagrange multiplier associated with the constraint
$\contwo_{ts}(x_s) = 0$.  We then consider the Lagrangian
\begin{multline}
\label{EqnLagOrig}
\Lag_{\eparambar, \treedistvec}(\taupar, \newlag) =
\inprod{\taupar}{\eparambar} + \\
\sum_{(s,t) \in \edge} \treedist_{st} \biggr [ \sum_{x_s\in
\statesp_s} \newlag_{ts}(x_s) \contwo_{ts}(x_s) + \sum_{x_t \in
\statesp_t} \newlag_{st}(x_t) \contwo_{st}(x_t) \biggr].
\end{multline}
Note that we have rescaled the Lagrange multipliers by the edge
appearance probabilities $\treedist_{st} > 0$ so as to make the
connection to the messages in Algorithm~\algtrw$\:$ as explicit as
possible.  For any vector of Lagrange multipliers $\newlag$, the dual
function is defined by the maximization \mbox{$\Qdual_{\eparambar;
\treedistvec}(\newlag) \defn \max_{\taupar \in \Normset}
\Lag_{\eparambar, \treedistvec}(\taupar, \newlag)$,} over the
constraint set
\begin{equation}
\label{EqnDefnConset}
\Normset \defn \{ \taupar \: | \; \taupar \geq 0, \; \sum_{x_s}
\taupar_{s}(x_s) = 1, \; \; \sum_{x_s, x_t} \taupar_{st}(x_s, x_t) = 1
\}.
\end{equation}

Now our goal is to specify how to choose a particular Lagrange
multiplier vector $\newlag^*$ in terms of the log messages
$\logmess^*$, or equivalently the pseudo-max-marginals
$\grmaxmarg^*_s$ and $\grmaxmarg^*_{st}$ defined by the messages
$\mess^* \defn \exp(\logmess^*)$.  To define the link between
$\newlag^*$ and $\grmaxmarg^*$, we let $r$ be an arbitrary node of the
graph, and suppose that every tree $\treegr \in \supp(\treedistvec)$
is rooted at $r$, and the remaining edges are directed from
parent-to-child.  More formally, for each node $s \neq r$, let
$\pa(s)$ denote its unique parent.  We now define
\begin{eqnarray}
\label{EqnNewLag}
\newlag^*_{ts}(x_s) & \defn & \logmess^*_{ts}(x_s) - \sum_{\{ \treegr
\, \big | \, \pa(s) = t\}} \treedist(\treegr) \log
\grmaxmarg^*_s(x_s).
\end{eqnarray}
With this definition, the Lagrangian evaluated at $\newlag^*$ takes a
particular form:
\begin{lemma}
\label{LemTech}
With $\newlag^*$ defined in equation~\eqref{EqnNewLag}, the
Lagrangian 
$\Lag_{\eparambar, \treedistvec}(\taupar, \newlag^*)$ can be
written as
\begin{multline*}
\sum_{\treegr} \treedist(\treegr) \Fcon(\taupar, \grmaxmarg^*;
\treegr)  \\
+ \sum_{s \in \vertex} \kappa_s \sum_{x_s} \taupar_s(x_s) +
\sum_{(s,t) \in \edge} \kappa_{st} \sum_{x_s, x_t} \taupar_{st}(x_s,
x_t),
\end{multline*}
where $\kappa_s$ and $\kappa_{st}$ are constants, and $\Fcon(\taupar,
\grmaxmarg^*; \treegr)$ is given by
\begin{multline}
\label{EqnDefnFcon}
\sum_{s \neq r} \sum_{(x_s, x_{\pa(s)})} \taupar_{st}(x_s, x_{\pa(s)})
\log \frac{\grmaxmarg^*_{s \pa(s)}(x_s,
x_{\pa(s)})}{\grmaxmarg^*_s(x_s)} \\
+ \sum_{x_r} \taupar_{r}(x_r) \log \grmaxmarg^*_{r}(x_r).
\end{multline}
\end{lemma}
\begin{proof}
See Appendix~\ref{AppTech}.
\end{proof} \hfill  \\
We now determine the form of the dual function $\Qdual_{\eparambar;
\treedistvec}(\newlag^*)$.  Note that $\sum_{x_s} \taupar_s(x_s) = 1$
and $\sum_{x_s, x_t} \taupar_{st}(x_s, x_t) = 1$ on the constraint set
$\Normset$ defined in equation~\eqref{EqnDefnConset}, so that the
terms involving $\kappa_s$ and $\kappa_{st}$ play no role in the
optimization.  Using the definition of $\Qdual$ and
Lemma~\ref{LemTech}, we write
\begin{eqnarray}
\label{EqnDualBound}
\Qdual_{\eparambar; \treedistvec}(\newlag^*) & = & \max_{\taupar \in
\Normset} \sum_{\treegr} \treedist(\treegr) \Fcon(\taupar,
\grmaxmarg^*; \treegr) + \kappa \nonumber \\
& \stackrel{(a)}{\leq} & \sum_{\treegr} \treedist(\treegr)
\max_{\taupar \in \Normset} \Fcon(\taupar, \grmaxmarg^*; \treegr) +
\kappa,
\end{eqnarray}
where $\kappa \defn \sum_s \kappa_s + \sum_{(s,t)} \kappa_{st}$.

Now since $\grmaxmarg^*$ satisfies the OS criterion by assumption, we
can find a vector $\xvec^*$ that achieves $\arg \max_{x_s}
\grmaxmarg^*(x_s)$ for every node, and $\arg \max_{x_s, x_t}
\grmaxmarg_{st}^*(x_s, x_t)$ for every edge.  Consider the
pseudomarginal vector given by
\begin{subequations}
\begin{eqnarray}
\label{EqnDefnTaustar}
\taupar^*_s(x_s) & \defn & \delta(x_s=x^*_s), \\
\taupar_{st}(x_s, x_t) & \defn & \delta(x_s = x^*_s) \delta(x_t = x^*_t).
\end{eqnarray}
\end{subequations}
The following lemma is proved in Appendix~\ref{AppFinal}:
\begin{lemma}
\label{LemFinal}
For each tree $\treegr$, we have $\max_{\taupar \in \Normset}
\Fcon(\taupar, \grmaxmarg^*; \treegr) = \Fcon(\taupar^*, \grmaxmarg^*;
\treegr)$.
\end{lemma}
This lemma shows that each of the maxima on the RHS of
equation~\eqref{EqnDualBound} are achieved at the same $\taupar^*$, so
that the inequality labeled (a) in equation~\eqref{EqnDualBound} in
fact holds with equality.  Consequently, we have
\begin{equation}
\label{EqnDualVal}
\Qdual_{\eparambar; \treedistvec}(\newlag^*) = \sum_{\treegr}
\treedist(\treegr) \Fcon(\taupar^*, \grmaxmarg^*; \treegr) + \kappa \;
= \; \Lag_{\eparambar, \treedistvec}(\taupar^*, \newlag^*).
\end{equation}
Since $\taupar^*$ by construction satisfies all of the marginalization
constraints (i.e., $\sum_{x_s} \taupar^*_{st}(x_s, x_t) =
\taupar^*_s(x_s)$), the Lagrangian reduces to the cost function, so
that we have shown that the dual value $\Qdual_{\eparambar;
\treedistvec}(\newlag^*)$ is equal to
\begin{equation*}
\sum_{s \in \vertex} \sum_{x_s} \taupar^*_s(x_s) \eparambar_s(x_s) +
\sum_{(s,t) \in \edge} \sum_{x_s, x_t} \taupar^*_{st}(x_s, x_t)
\eparambar_{st}(x_s, x_t),
\end{equation*}
or equivalently by $\sum_{s \in \vertex} \eparambar_s(x^*_s) +
\sum_{(s,t) \in \edge} \eparambar_{st}(x^*_s, x^*_t)$, which is the
optimal primal value.  By strong duality, the pair $(\taupar^*,
\newlag^*)$ are primal-dual optimal.

\end{proof} \hfill  \\
For a general graph with cycles, the above proof does not establish
that any fixed point of Algorithm~\algtrw$\:$ (i.e., one for which
$\grmaxmarg^*$ does not satisfy the OS criterion) necessarily
specifies a dual-optimal solution of the LP relaxation.  Indeed, in
follow-up work, Kolmogorov~\cite{Kol05b} has constructed a particular
fixed point, for which the OS criterion is \emph{not} satisfied, that
does not specify an optimal dual solution.  However, for problems that
involve only binary variables and pairwise interactions, Kolmogorov
and Wainwright~\cite{KolWai05} have strengthened
Theorem~\ref{ThmExactMAP} to show that a message-passing fixed point
always specifies an optimal dual solution.

However, on a tree-structured graph, the tree-reweighted max-product
updates reduce to the ordinary max-product (min-sum) updates, and any
fixed point $\grmaxmarg^*$ must satisfy the OS criterion.  In this
case, we can use Theorem~\ref{ThmExactMAP} to obtain the following
corollary:
\begin{corollary}[Ordinary max-product]
\label{CorOrdMax}
For a tree-structured graph $\treegr$, the ordinary max-product
algorithm is an iterative method for solving the dual of the exact LP
representation of the MAP problem:
\begin{eqnarray}
\label{EqnTreeMAPLP}
\max_{\xvec \in \stsp} \inprod{\eparambar}{\clipotvec(\xvec)} & = &
\max_{\taupar \in \Locset(\treegr) } \inprod{\eparambar}{\taupar}.
\end{eqnarray}
\end{corollary}
\begin{proof}
By Lemma~\ref{LemMargRep}, the MAP problem $\max_{\xvec \in \stsp}
\inprod{\eparambar}{\clipotvec(\xvec)}$ has the alternative LP
representation as $\max_{\taupar \in \Margset(\treegr)}
\inprod{\eparambar}{\taupar}$.  By Corollary~\ref{CorTreeExact}, the
relaxation based on $\Locset(\treegr)$ is exact for a tree, so that
$\max_{\taupar \in \Margset(\treegr)} \inprod{\eparambar}{\taupar} =
\max_{\taupar \in \Locset(\treegr)} \inprod{\eparambar}{\taupar}$,
from which equation~\eqref{EqnTreeMAPLP} follows.  For the case of a
tree, the only valid choice of $\trdistvec$ is the vector of all ones,
so the tree-reweighted updates must be equivalent to the ordinary
max-product algorithm.  The result then follows from
Theorem~\ref{ThmExactMAP}.
\end{proof} \hfill  \\

\subsection{Additional properties}

As proved in Corollaries~\ref{CorTreeExact} and~\ref{CorOrdMax}, the
techniques given here are always exact for tree-structured graphs.
For graphs with cycles, the general mode-finding problem considered
here includes many NP-hard problems, so that our methods cannot be
expected to work for all problems.  In general, their
performance---more specifically, whether or not a MAP configuration
can be obtained----depends on both the graph structure, and the form
of the parameter vector $\eparambar$.  In parallel and follow-up work
to this paper, we have obtained more precise performance guarantees
for our techniques when applied to particular classes of problems
(e.g., binary linear coding problems~\cite{Feldman05}; binary
quadratic programs~\cite{KolWai05}).  We discuss these results in more
detail in the sequel.

In this section, we begin with a comparison of reweighted max-product
to the standard max-product.  In particular, we explicitly construct a
simple problem for which the ordinary max-product algorithm outputs an
incorrect answer, but for which the reweighted updates provably find
the global optimum. We then demonstrate the properties of edge-based
versus tree-based updates, and discuss their relative merits.

\subsubsection{Comparison with ordinary max-product} 

Recall that for any graph with cycles, the tree-reweighted max-product
algorithm (Algorithm~\algtrw) differs from the ordinary
max-product algorithm in terms of the reweighting of messages and
potential functions, and the involvement of the reverse direction
messages.  Here we illustrate with a simple example that these
modifications are in general necessary for a message-passing algorithm
to satisfy the exactness condition of Theorem~\ref{ThmExactMAP}(a).
More precisely, we construct a fixed point of the ordinary max-product
algorithm that satisfies the optimum-specification (OS) criterion, yet
the associated configuration $\mxvec$ is not MAP-optimal.

\newcommand{\parone}{\ensuremath{\alpha}}
\newcommand{\partwo}{\ensuremath{\beta}}
\newcommand{\edgepar}{\ensuremath{\gamma}}
\newcommand{\pmax}{\ensuremath{\nu^*}}
\newcommand{\graphdia}{\ensuremath{\graph_{\operatorname{dia}}}}

\begin{figure}[h]
\bec
\begin{tabular}{c}
\psfrag{#1#}{$1$}
\psfrag{#2#}{$2$}
\psfrag{#3#}{$3$}
\psfrag{#4#}{$4$}
\widgraph{.20\textwidth}{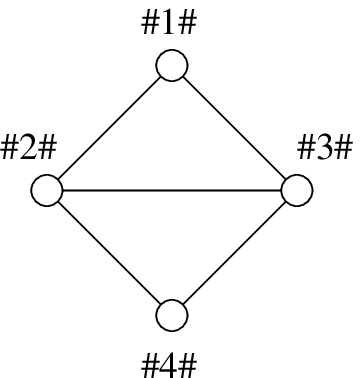}
\end{tabular}
\enc
\caption{Simple diamond graph $\graphdia$. }
\label{FigGraphdia}
\end{figure}

In particular, consider the simple graph $\graphdia$ shown in
Figure~\ref{FigGraphdia}, and suppose that we wish to maximize a cost
function of the form 
\begin{equation}
\label{EqnDiaCost}
\parone (x_1 + x_4) + \partwo (x_2 + x_3) + \edgepar \sum_{(s,t) \in
\edge} \big[ x_s (x_t-1) + x_t (x_s - 1) \big]
\end{equation}
Here the minimization is over all binary vectors $\xvec \in
\{0,1\}^4$, and $\parone, \partwo$ and $\edgepar$ are parameters to be
specified.  By design, the cost function~\eqref{EqnDiaCost} is such
that if we make $\edgepar$ sufficiently positive, then any optimal
solution will either be $0^4 \defn
\begin{bmatrix} 0 & 0 & 0 & 0
\end{bmatrix}$ or $1^4 \defn \begin{bmatrix} 1 & 1 & 1 & 1 \end{bmatrix}$.
More concretely, suppose that we set $\parone = 0.31$, $\partwo =
-0.30$, and $\edgepar = 2.00$.  With these parameter settings, it is
straightforward to verify that the optimal solution is $1^4$.
However, if we run the \emph{ordinary} max-product algorithm on this
problem, it converges to a set of singleton and edge-based
pseudo-max-marginals $\pmax$ of the form:
\begin{eqnarray*}
\pmax_s(x_s) & = & \begin{cases}
\begin{bmatrix} 1 & 0.0250
\end{bmatrix}  \quad \mbox{for $s \in \{1,4\}$} \\
\begin{bmatrix} 1 & 0.0034 \end{bmatrix} \quad \mbox{if $s \in \{2,3\}$.} 
	       \end{cases} \\
\pmax_{st}(x_s,x_t) & = & \begin{cases}
\begin{bmatrix} 1 & 0.0034 \\ 0.0034 & 0.0006 
\end{bmatrix} \quad \mbox{for $(s,t) = (2,3)$}  \\
& \\
\begin{bmatrix} 1 & 0.0250 \\ 0.0025 & 0.0034 \end{bmatrix} \quad \mbox{otherwise.} 
	       \end{cases}
\end{eqnarray*}
Note that these pseudo-max-marginals and the configuration $0^4$
satisfy the OS criterion (since the optimum is uniquely attained at
$x_s = 0$ for every node, and at the pair $(x_s, x_t) = (0,0)$ for
every edge); however, the global configuration $0^4$ is \emph{not} the
MAP configuration for the original problem.  This problem shows that
the ordinary max-product algorithm does not satisfy an exactness
guarantee of the form given in Theorem~\ref{ThmExactMAP}.

In fact, for the particular class of problems exemplified by
equation~\eqref{EqnDiaCost}, we can make a stronger assertion about
the tree-reweighted max-product algorithm: namely, it will never fail
on a problem of this general form, where the couplings $\gamma$ are
non-negative.  More specifically, in follow-up work to the current
paper, Kolmogorov and Wainwright~\cite{KolWai05} have established
theoretical guarantees on the performance of tree-reweighted
message-passing for problems with binary variables and pairwise
couplings.  First, it can be shown that TRW message-passing always
succeeds for any submodular binary problem, of which the example given
in Figure~\ref{FigGraphdia} is a special case.  Although it is
known~\cite{Greig89} that such problems can be solved in polynomial
time via reduction to a max-flow problem, it is nonetheless
interesting that tree-reweighted message-passing is also successful
for this class of problems.  An additional result~\cite{KolWai05} is
that for any pairwise binary problem (regardless of nature of the
pairwise couplings), any variable $s$ that is uniquely specified by a
TRW fixed point (i.e., for which the set $\arg \max_{x_s}
\grmaxmarg^*_s(x_s)$ is a singleton) is guaranteed to be correct in
some globally optimal configuration.  Thus, TRW fixed points can
provide useful information about parts of the MAP optimal solution
even when the OS criterion is not satisfied.

\subsubsection{Comparison of edge-based and tree-based updates}
\label{SecAdd}
In this section, we illustrate the empirical performance of the
edge-based and tree-based updates on some sample problems.  So as to
allow comparison to the optimal answer even for large problems, we
focus on binary problems.  In this case, the theoretical
guarantees~\cite{KolWai05} described above allow us to conclude that
the tree-reweighted method yields correct information about (at least
part of the) optimum, without any need to compute the exact MAP
optimum by brute force.  Thus, we can run simply the TRW
algorithm---either the edge-based or tree-based updates---on any
submodular binary problem, and be guaranteed that given a fixed point,
it will either specify a globally optimal configuration (for
attractive couplings), or that any uniquely specified variables (i.e.,
for which $\arg \max_{x_s} \grmaxmarg^*_s(x_s)$ is a singleton) will
be correct (for arbitrary binary problems).
\begin{figure}[h]
\begin{center}
\begin{tabular}{c}
\widgraph{.35\textwidth}{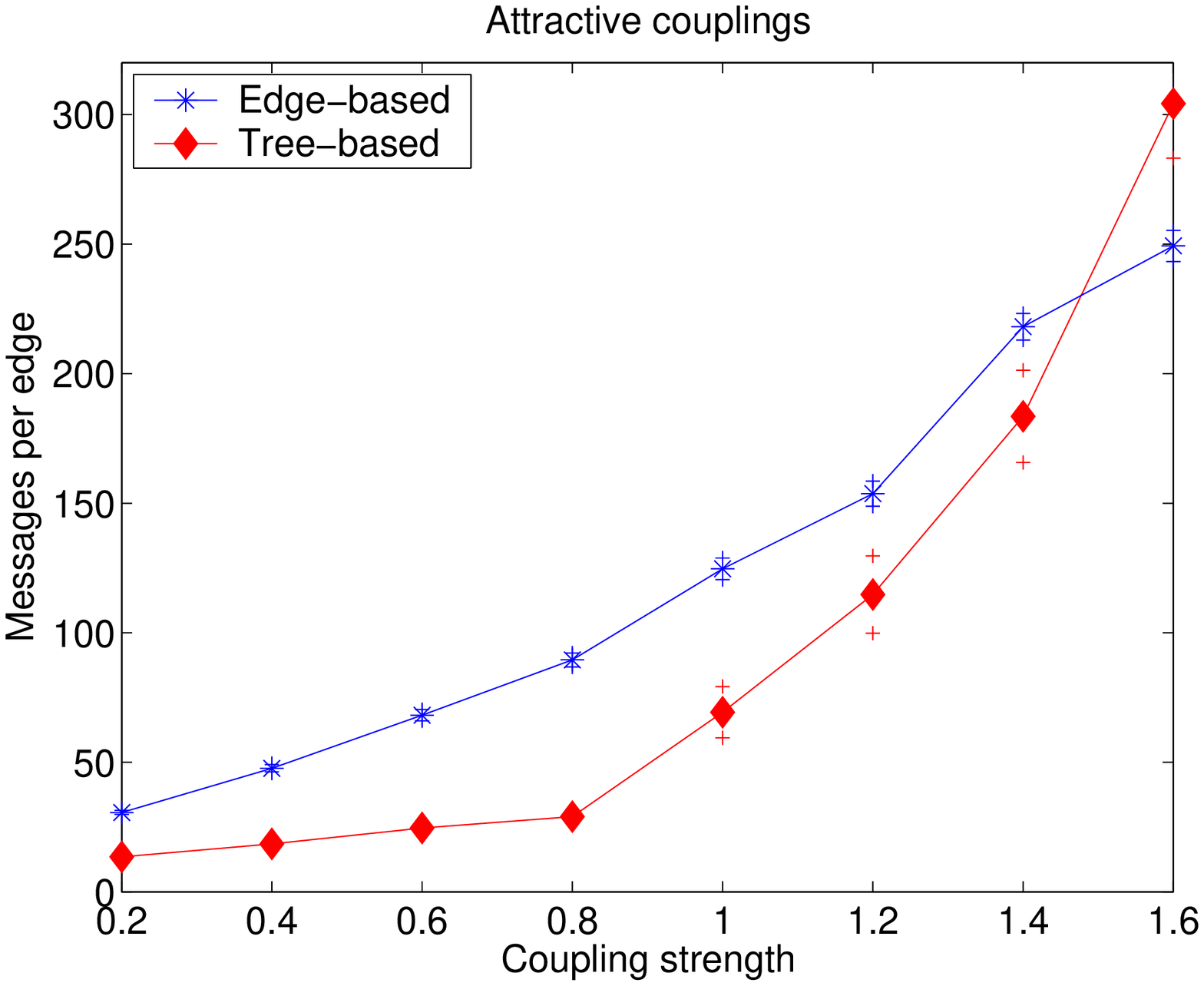} \\
(a) \\
\widgraph{.35\textwidth}{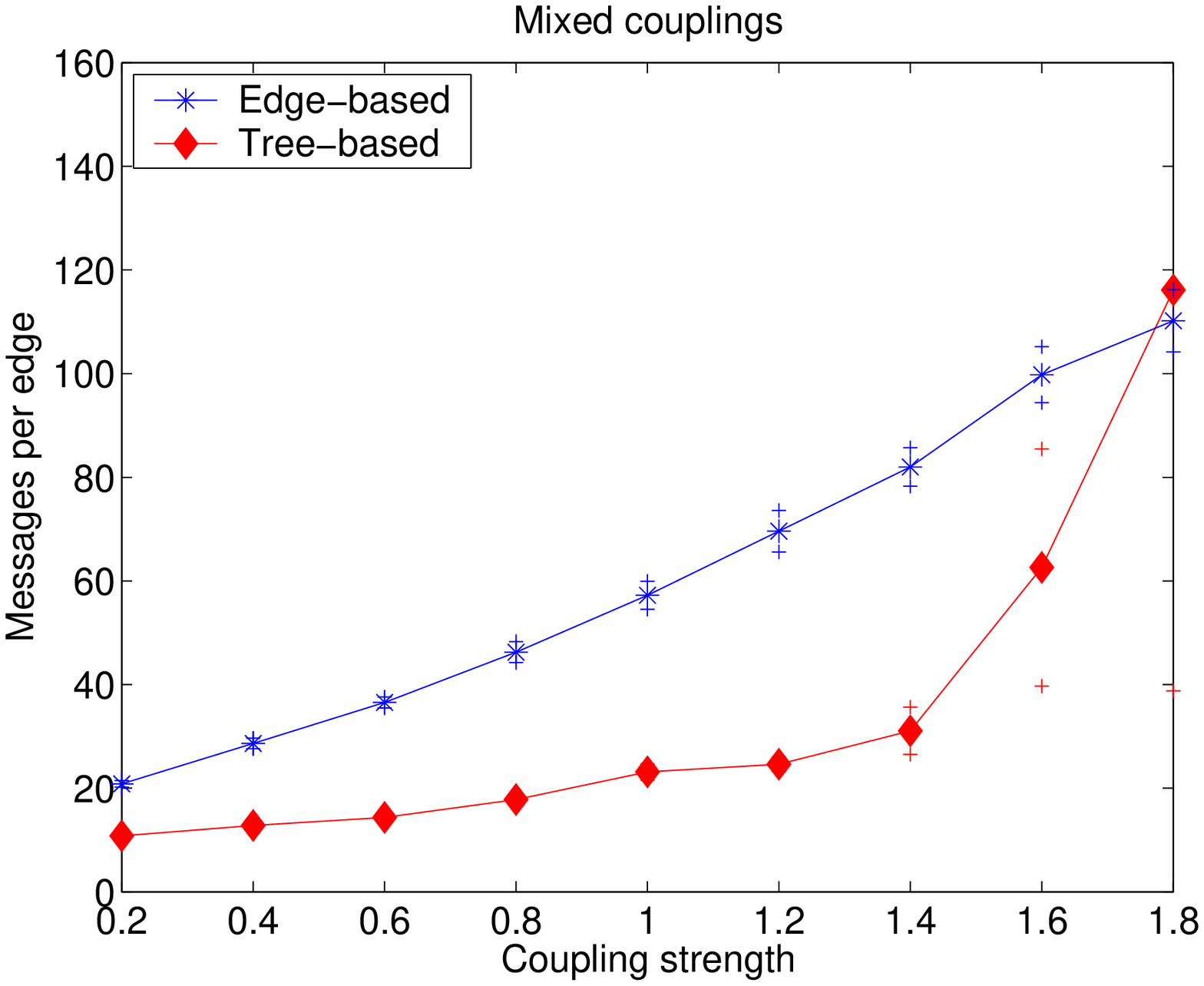} \\
(b)
\end{tabular}
\caption{Comparison of parallel edge-based message-passing
(Algorithm~\algtrw) and tree-based updates (Algorithm~\algtree$\:$
described in Appendix~\ref{AppTreeBase}) on a nearest-neighbor grid
with $\spnodenum = 400$ variables. (a) Attractive couplings.  (b)
Mixed couplings.}
\label{FigTimes}
\end{center}
\end{figure}

Our comparison is between the parallel edge-based form of reweighted
message-passing (Algorithm~\algtrw), and the tree-based
Algorithm~\algtree$\;$ described in Appendix~\ref{AppTreeBase}.  We focus
on the amount of computation, as measured by the number of messages
passed along each edge, required to \emph{either} compute the fixed
point (up to $\eps = 1\times10^{-8}$ accuracy), \emph{or}---in the
case of tree-based updates---to find a configuration $\xvec^*$ on
which all trees agree.  In this latter case,
Proposition~\ref{PropTightBound} guarantees that the shared
configuration $\xvec^*$ must be globally MAP-optimal for the original
problem, so that there is no need to perform any further
message-passing.

\newcommand{\Uni}{\mathcal{U}}

We performed trials on problems in the Ising form~\eqref{EqnIsing},
defined on grids with $\spnodenum = 400$ nodes.  For the edge-based
updates, we used the uniform setting of edge appearance probabilities
$\treedist_{st} = \frac{\spnodenum-1}{|\edge|}$; for the tree-based
updates, we used two spanning trees, one with the horizontal rows plus
an connecting column and the rotated version of this tree, placing
weight $\treedist(\treegr^i) = \frac{1}{2}$ on each tree $i=1,2$.  In
each trial, the single node potentials were chosen randomly as
$\eparam_s \sim \Uni[-1,1]$, whereas the edge couplings were chosen in
one of the following two ways.  In the attractive case, we chose the
couplings as $\eparam_{st} \sim \Uni[0, \gamma]$, where $\gamma \geq
0$ is the edge strength.  In the mixed case, we chose $\eparam_{st}
\sim \Uni[-\frac{\gamma}{2}, \frac{\gamma}{2}]$.  In both cases, we
used damped forms of the updates (linearly combining messages or
pseudo-max-marginals in the logarithmic domain) with damping parameter
$\lambda = 0.50$.

We investigated the algorithmic performance for a range of coupling
strengths $\gamma$ for both attractive and mixed cases.  For the
attractive case, the TRW algorithm is theoretically
guaranteed~\cite{KolWai05} to always find an optimal MAP
configuration.  For the mixed case, the average fraction of variables
in the MAP optimal solution that the reweighted message-passing
recovered was above $80\%$ for all the examples that we considered;
for mixed problems with weaker observation terms $\eparam_s$, this
fraction can be lower~\cite[see]{KolWai05}.  Figure~\ref{FigTimes}
shows results comparing the behavior of the edge-based and tree-based
updates.  In each panel, plotted on the $y$-axis is the number of
messages passed per edge (before achieving the stopping criterion)
versus the coupling strength $\gamma$.  Note that for more weakly
coupled problems, the tree-based updates consistently find the MAP
optimum with lower computation than the edge-based updates.  As the
coupling strength is increased, however, the performance of the
tree-based updates slows down and ultimately becomes worse than the
edge-based updates.  In fact, for strong enough couplings, we observed
on occasion that the tree-based updates could fail to converge, but
instead oscillate (even with small damping parameters).  These
empirical observations are consistent with subsequent observations and
results by Kolmogorov~\cite{Kol05}, who developed a modified form of
tree-based updates for which certain convergence properties are
guaranteed.  (In particular, in contrast to the tree-based schedule
given in Appendix~\ref{AppTreeBase}, they are guaranteed to generate a
monotonically non-increasing sequence of upper bounds.)

\subsubsection{Related work}
\label{SecApply}

In related work involving the methods described here, we have found
several applications in which the tree relaxation and iterative
algorithms described here are useful.  For instance, we have applied
the tree-reweighted max-product algorithm to a distributed data
association problem involving multiple targets and
sensors~\cite{Chen03}.  For the class of problem considered, the
tree-reweighted max-product algorithm converges, typically quite
rapidly, to a provably MAP-optimal data association.  In other
colloborative work, we have also applied these methods to decoding
turbo-like and low density parity check (LDPC)
codes~\cite{Feldman02aller,Feldman03,Feldman05}, and provided
finite-length performance guarantees for particular codes and
channels.  In the context of decoding, the fractional vertices of the
polytope $\Locset(\graph)$ have a very concrete interpretation as
pseudocodewords~\cite[e.g.,]{Forney01c,Frey01b,Horn_phd,Wiberg_phd}.
More broadly, it remains to further explore and analyze the range of
problems for which the iterative algorithms and LP relaxations
described here are suitable.

\section{Discussion}

\label{SecDiscuss}

In this paper, we demonstrated the utility of convex combinations of
tree-structured distributions in upper bounding the value of the
maximum a posteriori (MAP) configuration on a Markov random field
(MRF) on a graph with cycles.  A key property is that this upper bound
is tight if and only if the collection of tree-structured
distributions shares a common optimum. Moreover, when the upper bound
is tight, then a MAP configuration can be obtained for the original
MRF on the graph with cycles simply by examining the optima of the
tree-structured distributions.  This observation motivated two
approaches for attempting to obtain tight upper bounds, and hence MAP
configurations.  First of all, we proved that the Lagrangian dual of
the problem is equivalent to a linear programming (LP) relaxation,
wherein the marginal polytope associated with the original MRF is
replaced with a looser constraint set formed by tree-based consistency
conditions.  Interestingly, this constraint set is equivalent to the
constraint set in the Bethe variational formulation of the sum-product
algorithm~\cite{Yedidia02}; in fact, the LP relaxation itself can be
obtained by taking a suitable limit of the ``convexified'' Bethe
variational problem analyzed in our previous
work~\cite{Wainwright02a,Wainwright05}.  Second, we developed a family of {\em
tree-reweighted max product} algorithms that reparameterize a
collection of tree-structured distributions in terms of a common set
of pseudo-max-marginals on the nodes and edges of the graph with
cycles. When it is possible to find a configuration that is locally
optimal with respect to every single node and edge
pseudo-max-marginal, then the upper bound is tight, and the MAP
configuration can be obtained.  Under this condition, we proved that
fixed points of these message-passing algorithms specify dual-optimal
solutions to the LP relaxation.  A corollary of this analysis is that
the ordinary max-product algorithm, when applied to trees, is solving
the Lagrangian dual of an exact LP formulation of the MAP estimation
problem.

Finally, in cases in which the methods described here do not yield MAP
configurations, it is natural to consider strengthening the relaxation
by forming clusters of random variables, as in the Kikuchi
approximations described by Yedidia et al.~\cite{Yedidia02}.  In the
context of this paper, this avenue amounts to taking convex
combinations of hypertrees, which (roughly speaking) correspond to
trees defined on clusters of nodes. Such convex combinations of
hypertrees lead, in the dual reformulation, to a hierarchy of
progressively tighter LP relaxations, ordered in terms of the size of
clusters used to form the hypertrees.  On the message-passing side, it
is also possible to develop hypertree-reweighted forms of
generalizations of the max-product algorithm.

\vtiny

\mypara{Acknowledgments}  We thank  Jon Feldman and David Karger for 
helpful discussions.
\appendix

\subsection{Conversion from factor graph to pairwise interactions}
\label{AppConv}

In this appendix, we briefly describe how any factor graph description
of a distribution over a discrete (multinomial) random vector can be
equivalently described in terms of a pairwise Markov random
field~\cite{FreemanMAP01}, to which the pairwise LP relaxation based
on $\Locset(\graph)$ specified by equations~\eqref{EqnDefnNormalizea},
~\eqref{EqnDefnNormalizeb} and~\eqref{EqnDefnMarginalize} can be
applied.  To illustrate the general principle, it suffices to show how
to convert a factor $f_{123}$ defined on the triplet $\{X_1, X_2, X_3
\}$ of random variables into a pairwise form.  Say that each $X_i$
takes values in some finite discrete space $\statesp_i$.

Given the factor graph description, we associate a new random variable
$Z$ with the factor node $f$, which takes values in the Cartesian
product space $\mathcal{Z} \defn \statesp_1 \times \statesp_2 \times
\statesp_3$.  In this way, each possible value $z$ of $Z$ can be put
in one-to-one correspondence with a triplet $(z_1, z_2, z_3)$, where
$z_i \in \statesp_i$.  For each $s \in \{1,2,3\}$, we define a
pairwise compatibility function $\compat_{fs}$, corresponding to the
interaction between $Z$ and $X_s$, by
\[
\compat_{fs}(z, x_s) \defn \Ind[z_s =x_s],
\]
where $\Ind[z_s = x_s]$ is a $\{0,1\}$-valued indicator function for
the event $\{z_s = x_s \}$.   We set the singleton compatility functions
as
\[
\compat_f(z) = f(z_1, z_2, z_3),  \quad \mbox{and} \quad \compat_s(x_s) = 1.
\]
With these definitions, it is straightforward to verify that the
augmented distribution given by
\begin{equation}
\label{EqnAugModel}
\compat_f(z) \; \prod_{s =1}^3 \compat_s(x_s) \; \prod_{s=1}^3
\compat_{fs}(z, x_s)
\end{equation}
marginalizes down to $f(x_1, x_2, x_3)$.  Thus, our augmented model
with purely pairwise interactions faithfully captures the interaction
among the triplet $\{x_1, x_2, x_3\}$.

Finally, it is straightforward to verify that if we apply the pairwise
LP relaxation based on $\Locset(\graph)$ to the augmented
model~\eqref{EqnAugModel}, it generates an LP relaxation in terms of
the $x_i$ variables that involves singleton pseudomarginal
distributions $\taupar_s$, and a pseudomarginal $\taupar_{N(f)}$ over
the variable neighborhood of each factor $f$.  These pseudomarginals
are required to be non-negative, normalized to one, and to satisfy the
pairwise consistency conditions
\begin{eqnarray}
\sum_{x_t, \; t \in N(f) \bk \{s \} } \taupar(x_{N(f)}) & = &  \taupar_s(x_s)
\end{eqnarray}
for all $s \in N(f)$, and for all factor nodes $f$.  When the factor
graph defines an LDPC code, this procedure generates the LP relaxation
studied in Feldman et al.~\cite{Feldman05}.  More generally, this LP
relaxation can be applied to factor graph distributions other than those
associated with LDPC codes.

\subsection{Proof of Lemma~\ref{LemIndicate}}
\label{AppIndicate}

By definition, we have $\Partinf(\myproj{\eparam}{\treegr}) \defn
\max_{\xvec \in \stsp}
\inprod{\myproj{\eparam}{\treegr}}{\clipotvec(\xvec)}$.  We re-write
this function in the following way:
\begin{eqnarray*}
\Partinf(\myproj{\eparam}{\treegr}) & \overset{(a)}{=} & \max_{\tausym
\in \Locset(\graph)} \inprod{\myproj{\eparam}{\treegr}}{\tausym} \\
& \overset{(b)}{=} & \max_{\tausym \in \Locset(\graph; \treegr) }
\inprod{\myproj{\eparam}{\treegr}}{\tausym}
\end{eqnarray*}
where equality (a) follows from Lemma~\ref{LemMargRep}, and equality
(b) follows because $\myproj{\eparam}{\treegr}_\sumind = 0$ for all
$\sumind \notin \iset(\treegr)$.  In this way, we recognize
$\Partinf(\myproj{\eparam}{\treegr})$ as the {\em support function} of
the set $\Locset(\graph; \treegr)$, from which it
follows~\cite{Hiriart1} that the conjugate dual is the indicator
function of $\Locset(\graph; \treegr)$, as specified in
equation~\eqref{EqnIndicate}.

For the sake of self-containment, we provide an explicit proof of this
duality relation here.  If $\tausym$ belongs to $\Locset(\graph;
\treegr)$, then \mbox{$\inprod{ \myproj{\eparam}{\treegr}}{\tausym} -
\Partinf(\myproj{\eparam}{\treegr}) \; \leq \; 0$} holds for all
$\myproj{\eparam}{\treegr} \in \neweflatman(\treegr)$, with equality
for $\myproj{\eparam}{\treegr} = 0$.  From this relation, we conclude
that
\[
\sup_{\myproj{\eparam}{\treegr} \in \neweflatman(\treegr)} \inprod{
\myproj{\eparam}{\treegr}}{\tausym} -
\Partinf(\myproj{\eparam}{\treegr}) = 0
\]
whenever $\tausym \in \Locset(\graph; \treegr)$.

On the other hand, if $\tausym \notin
\Locset(\graph; \treegr)$, then by the (strong) separating hyperplane
theorem~\cite{Hiriart1}, there must exist some vector $\gamma$ and
constant $\beta$ such that (i) $ \;
\inprod{\gamma}{\meanpar} \leq \beta$ for all
$\meanpar \in \Locset(\graph; \treegr)$; and (ii) $
\; \inprod{\gamma}{\tausym} > \beta$.  Since
conditions (i) and (ii) do not depend on elements $\gamma_\sumind$
with $\sumind \notin \iset(\treegr)$, we can take \mbox{$\gamma \equiv
\myproj{\gamma}{\treegr} \in \neweflatman(\treegr)$} without loss of
generality.  We then have
\begin{equation}
\label{EqnScaleQuan}
  \inprod{\myproj{\gamma}{\treegr}}{\tausym} -
  \Partinf(\myproj{\gamma}{\treegr}) \; \geq \;
  \inprod{\myproj{\gamma}{\treegr}}{\tausym} -
  \beta \: > \; 0.
\end{equation}
Note that conditions (i) and (ii) are preserved under scaling of both
$\gamma(\treegr)$ and $\beta$ by a positive number, so that can send
the quantity~\eqref{EqnScaleQuan} to positive infinity.  We thus
conclude that
\[
\sup_{\eparam(\treegr) \in \neweflatman(\treegr)} \big \{ \inprod{
\myproj{\eparam}{\treegr}}{\tausym} -
\Partinf(\myproj{\eparam}{\treegr}) \big \} \; = \;  +\infty
\]
whenever $\tausym \notin \Locset(\graph; \treegr)$.  This completes
the proof of the lemma.

\subsection{Tree-based updates}
\label{AppTreeBase}

This appendix provides a detailed description of tree-based updates.
In this scheme, each iteration involves multiple rounds of
message-passing on each tree $\treegr$ in the support of
$\treedistvec$.  More specifically, the computational engine used
within each iteration is the ordinary max-product algorithm, applied
as an {\em exact} technique to compute max-marginals for each
tree-structured distribution.

At any iteration $n$, we let $\eparam^n(\treegr)$ denote a set of
exponential parameters for the tree $\treegr$.  To be clear, the
notation $\eparam^n(\treegr)$ used in this appendix differs slightly
from its use in the main text.  In particular, unlike in the main
text, we can have $\eparam^n_\sumind(\treegr) \neq
\eparam^n_\sumind(\treegr')$ for distinct trees $\treegr = \treegr'$
at immediate iterations, although upon convergence this equality will
hold.  Each step of the algorithm will involve computing, for every
tree $\treegr \in \supp(\treedistvec)$, the max-marginals
$\grmaxmarg^n(\treegr)$, associated with the tree-structured
distribution $p(\xvec; \eparam(\treegr))$.  (Once again, unlike the
main text, we need not have $\grmaxmarg_\sumind^n(\treegr) =
\grmaxmarg_\sumind^n(\treegr')$ for distinct trees $\treegr =
\treegr'$.)  Overall, the tree-based updates take the form given in
Figure~\ref{FigAlgTreeBase}.

\bec
\begin{figure*}[t]
\framebox[0.98\textwidth]{\parbox{.95\textwidth}{
%
\noindent {\bf{Algorithm~\algtree: Tree-based updates}} \hfill \\
\begin{enumerate}
\item For each spanning tree $\treegr \in \supp(\treedistvec)$,
initialize $\eparam^0(\treegr)$ via
\begin{eqnarray*}
\eparam^0_s(\treegr) & = & \eparambar_s \; \; \forall \; \; s \in
\vertex, \\
\eparam^0_{st}(\treegr) & = & \frac{1}{\treedist_{st}}
\eparambar_\sumind \; \; \forall \; \; (s,t) \in \edge(\treegr),
\qquad \qquad \quad \eparam^0_{st}(\treegr) \; = \; 0 \; \; \forall \;
\; (s,t) \in \edge\bk\edge(\treegr).
\end{eqnarray*}
\item For iterations $n = 0,1,2,\ldots$, do the following: 
\ben
\item[(a)] For each tree $\treegr \in \supp(\treedistvec)$, apply the
ordinary max-product algorithm to compute the max-marginals
$\grmaxmarg^n(\treegr)$ corresponding to the tree-structured
distribution $p(\xvec; \eparam^n(\treegr))$.
\item[(b)] Check if the tree distributions share a common optimizing
configuration (i.e., if $\cap_{\treegr} \xoptset(\eparam^n(\treegr))$
is non-empty).   \\
\noindent (i) If yes, output any shared configuration and
terminate. \\
\noindent (ii) If not, check to see whether or not the following
agreement condition holds:
\begin{subequations}
\label{EqnMaxAgree}
\begin{eqnarray}
\label{EqnMaxAgreea}
\grmaxmarg_s(\treegr) & = & \grmaxmarg_s(\treegr') \qquad \forall \; s
\in \vertex, \qquad \forall \treegr, \treegr' \in
\supp(\treedistvec),\\
\label{EqnMaxAgreeb}
\grmaxmarg_{st}(\treegr) & = & \grmaxmarg_{st}(\treegr') \qquad
\forall \; \treegr, \; \treegr' \; \in \; \supp(\treedistvec) \;
\suchthat \; (s,t) \in \edge(\treegr) \cap \edge(\treegr').
\end{eqnarray}
\end{subequations}
If this agreement of all max-marginals holds, then terminate.
Otherwise, form a new exponential parameter $\eparamtil$ as follows:
\begin{subequations}
\label{EqnDefnNewe}
\begin{eqnarray}
\label{EqnDefnNewea}
\eparamtil^{n+1}_s & = & \sum_{\treegr} \treedist(\treegr) \log
\grmaxmarg^n_s(\treegr) \; \; \forall \; \; s \in \vertex \\
\label{EqnDefnNeweb}
\eparamtil^{n+1}_{st} & = & \sum_{\treegr \ni (s,t)}
\treedist(\treegr) \log
\frac{\grmaxmarg^n_{st}(\treegr)}{\grmaxmarg^n_s(\treegr)
\grmaxmarg^n_t(\treegr)} \; \; \forall \; \; (s,t) \in \edge
\end{eqnarray}
\end{subequations}
Define $\eparam^{n+1}(\treegr)$ on each tree $\treegr \in
\supp(\treedistvec)$ as in Step 1 with $\eparambar \equiv \eparamtil$,
and proceed to Step 2(a).
\een
\end{enumerate}
}}
\caption{Tree-based updates for finding a tree-consistent set of
pseudo-max-marginals.}
\label{FigAlgTreeBase}
\end{figure*}
\enc

\vtiny

\mypara{Termination} Observe that there are two possible ways in which
Algorithm~\algtree$\:$ can terminate.  On one hand, the algorithm
stops if in Step 2(b)(i), a collection of tree-structured
distributions is found that all share a common optimizing
configuration.  Herein lies the possibility of {\em finite
termination}, since there is no need to wait until the values of the
tree max-marginals $\grmaxmarg^n(\treegr)$ all agree for every tree.
Otherwise, the algorithm terminates in Step 2(b)(ii) if the
max-marginals for each tree all agree with one another, as stipulated
by equation~\eqref{EqnMaxAgree}.

\vtiny

A key property of the updates in Algorithm~\algtree$\:$ is that
they satisfy the $\treedistvec$-reparameterization
condition:
\begin{lemma}
\label{LemAdmissibleTree}
For any iteration $n$, the tree-structured parameters
$\{\eparam^n(\treegr)\}$ of Algorithm~\algtree$\:$ satisfy
$\sum_{\treegr} \treedist(\treegr) \eparam^n(\treegr) = \eparambar$.
\end{lemma}
\begin{proof}
The claim for $n=0$ follows from directly the initialization in Step
1.  In particular, we clearly have $\sum_{\treegr} \treedist(\treegr)
\eparam^0_s(\treegr) = \eparambar_s$ for any node $s \in \vertex$.
For any edge $(s,t) \in \edge$, we compute:
\begin{equation*}
\sum_{\treegr} \treedist(\treegr) \eparam^0_{st}(\treegr) \; = \;
\sum_{\treegr} \treedist(\treegr) \big[\frac{1}{\treedist_{st}}
\eparambar_{st}] \; = \; \eparambar_{st}.
\end{equation*}
To establish the claim for $n+1$, we proceed by induction.  By the
claim just proved, it suffices to show that $\eparamtil$, as defined
in equation~\eqref{EqnDefnNewe}, defines the same distribution as
$\eparambar$.

We begin by writing $\inprod{\eparamtil}{\clipotvec(\xvec)}$ as
\begin{equation*}
\sum_{s \in \vertex}
\sum_{x_s} \eparamtil_s(x_s) + \sum_{(s,t) \in \edge} \sum_{x_s, x_t}
\eparamtil_{st}(x_s, x_t)
\end{equation*}
Using the definition~\eqref{EqnDefnNewe}, we can re-express it as
follows
\begin{multline}
\label{EqnFinal}
 \sum_{\treegr} \treedist(\treegr) \Biggr \{ \sum_{s \in \vertex}
\sum_{x_s} \log \grmaxmarg^n_s(\treegr)(x_s) \\
+ \sum_{(s,t) \in \edge(\treegr)} \sum_{x_s, x_t} \log
\frac{\grmaxmarg^n_{st}(\treegr)(x_s,
x_t)}{\grmaxmarg^n_s(\treegr)(x_s) \grmaxmarg^n_t(\treegr)(x_t)}
\Biggr \}
\end{multline}
Recall that for each tree $\treegr$, the quantities
$\grmaxmarg^n(\treegr)$ are the max-marginals associated with the
distribution $p(\xvec; \eparam(\treegr))$.  Using the
fact~\eqref{EqnMaxMargRep} that the max-marginals specify a
reparameterization of $p(\xvec; \eparam^n(\treegr))$, each term within
curly braces is simply equal (up to an additive constant independent
of $\xvec$) to $\inprod{\eparam^n(\treegr)}{\clipotvec(\xvec)}$.
Therefore, by the induction hypothesis, the RHS of
equation~\eqref{EqnFinal} is equal to
$\inprod{\eparambar}{\clipotvec(\xvec)}$, so that the claim of the
lemma follows.
\end{proof} \hfill  \\

On the basis of Lemma~\ref{LemAdmissibleTree}, it is straightforward
to prove the analog of Theorem~\ref{ThmExactMAP}(a) for
Algorithm~\algtree.  More specifically, whenever it outputs a
configuration, it must be an exact MAP configuration for the original
problem $p(\xvec; \eparambar)$.

\subsection{Proof of Lemma~\ref{LemTech}}
\label{AppTech}

We first prove an intermediate result:
\begin{lemma}
\label{LemInterA}
Let $\grmaxmarg^*_s$ and $\grmaxmarg^*_{st}$ represent
pseudo-max-marginals, as defined as in equation~\eqref{EqnDefnPmax},
with the message $\mess^* \defn \exp(\logmess^*)$, where the
exponential is taken element-wise. Then the Lagrangian
$\Lag_{\eparambar, \treedistvec}(\taupar, \logmess^*)$ can be
written as
\[
\sum_{\treegr} \treedist(\treegr) \Gcon(\taupar, \logmess^*; \treegr)
+ \sum_s \kappa_s \sum_{x_s} \taupar_s(x_s) + \sum_{(s,t)} \kappa_{st}
\sum_{x_s, x_t} \taupar_{st}(x_s, x_t)
\]
where $\kappa_s$ and $\kappa_{st}$ are constants, and $\Gcon(\taupar,
\logmess^*; \treegr)$ is given by
\begin{multline*}
\sum_{s \in \vertex} \sum_{x_s} \taupar_{s}(x_s) \log
\grmaxmarg^*_{r}(x_r) \; + \; \\
\sum_{(s, t) \in \edge(\treegr)}
\sum_{(x_s, x_t)} \taupar_{st}(x_s, x_t) \log
\frac{\grmaxmarg^*_{st}(x_s, x_t)}{\grmaxmarg^*_s(x_s)
\grmaxmarg^*_t(x_t)}.
\end{multline*}
\end{lemma}
\begin{proof}
Straightforward algebraic manipulation allows us to re-express the
Lagrangian $\Lag_{\eparambar, \treedistvec}(\taupar, \lagmul)$ as
\begin{multline*}
\sum_{\treegr} \treedist(\treegr) \; \Biggr \{ \sum_{s \in \vertex}
\sum_{x_s} \taupar_s(x_s) \big[ \eparambar_s(x_s) + \sum_{t \in
\neigh(s)} \treedist_{st} \logmess^*_{ts}(x_s) \big] \\
+ \sum_{(s,t) \in \edge(\treegr)} \Big [ \sum_{x_s, x_t}
\taupar_{st}(x_s, x_t) \big[\frac{\eparambar_{st}(x_s,
x_t)}{\treedist_{st}} - \logmess^*_{st}(x_t) - \logmess^*_{ts}(x_s)
\big] \Big ] \Biggr \}.
\end{multline*}
Using the definition of $\grmaxmarg^*$ in terms of $\mess^* =
\exp(\logmess^*)$, we can then write $\Lag_{\eparambar,
\treedistvec}(\taupar, \logmess^*)$ in terms of $\Gcon(\taupar,
\logmess^*; \treegr)$ as
\begin{multline*}
\sum_{\treegr} \treedist(\treegr) \Gcon(\taupar, \logmess^*; \treegr)
+ \sum_s \kappa_s \sum_{x_s} \taupar_s(x_s) + \\
\sum_{(s,t)} \kappa_{st}
\sum_{x_s, x_t} \taupar_{st}(x_s, x_t),
\end{multline*}
where the constants $\kappa_s$ and $\kappa_{st}$ arise from the
normalization of $\grmaxmarg^*_s$ and $\grmaxmarg^*_{st}$.
\end{proof} \hfill  \\

Recall that we root all trees $\treegr$ a fixed vertex $r \in
\vertex$, so that each vertex $s \neq r$ has a unique parent denoted
$\pa(s)$.  Using the parent-to-child representation of the tree
$\treegr$, we can re-express $\Gcon(\taupar, \logmess^*; \treegr)$ as
\begin{multline*}
\Big \{\sum_{x_r} \taupar_r(x_r) \log \grmaxmarg^*_r(x_r) + \\
\sum_{(s, \pa(s))} \sum_{(x_s, x_{\pa(s)})} \taupar_{st}(x_s, x_t)
\log \frac{\grmaxmarg^*_{s\pa(s)}(x_s,
x_{\pa(s)})}{\grmaxmarg^*_{\pa(s)}(x_{\pa(s)})} \Big \} + \\
\sum_{s \neq r} \sum_{x_s} \log \grmaxmarg^*_s(x_s) \big[
\taupar_s(x_s) - \sum_{x_t} \taupar_{st}(x_s, x_t) \big]
\end{multline*}
Recalling the definition of $\Fcon_{\eparambar; \treedistvec}(\taupar,
\grmaxmarg; \treegr)$ from equation~\eqref{EqnDefnFcon}, we can
write $\sum_{\treegr} \treedist(\treegr) \Gcon(\taupar; \logmess^*; \treegr)$
as
\begin{multline*}
\sum_{\treegr} \treedist(\treegr) \Fcon_{\eparambar;
\treedistvec}(\taupar, \grmaxmarg^*; \treegr) + 
\\
\sum_{\treegr}
\treedist(\treegr) \sum_{s \neq r} \sum_{x_s} \log \grmaxmarg^*_s(x_s)
\big[ \taupar_s(x_s) - \sum_{x_t} \taupar_{st}(x_s, x_t) \big],
\end{multline*}
or equivalently as
\begin{multline*}
\sum_{\treegr} \treedist(\treegr) \Fcon_{\eparambar;
\treedistvec}(\taupar, \grmaxmarg^*; \treegr) + \\
\sum_{(s,t) \in \edge} \sum_{x_s} \big( \sum_{\{\treegr \, | t =
\pa(s)\}} \treedist(\treegr) \big) \log \grmaxmarg^*_s(x_s)
\contwo_{ts}(x_s),
\end{multline*}
where $\contwo_{ts}(x_s) = \taupar_s(x_s) - \sum_{x_t}
\taupar_{st}(x_s, x_t)$ is the constraint.  Note that for each fixed
$(s,t)$ and $x_s$ the term $\sum_{\{\treegr \, | t = \pa(s)\}}
\treedist(\treegr) \log \grmaxmarg^*_s(x_s)$ can be interpreted as a
contribution to the Lagrange multiplier associated with the constraint
$\contwo_{ts}(x_s)$.

Finally, since the Lagrangian is linear in the Lagrange multipliers,
we can use Lemma~\ref{LemInterA} to express the Lagrangian
$\Lag_{\eparambar, \treedistvec}(\taupar, \newlag^*; \treegr)$
as
\begin{multline*}
\sum_{\treegr} \treedist(\treegr) \Fcon_{\eparambar;
\treedistvec}(\taupar, \grmaxmarg^*; \treegr) + \sum_s \kappa_s
\sum_{x_s} \taupar_s(x_s) + \\ \sum_{(s,t)} \kappa_{st} \sum_{x_s,
x_t} \taupar_{st}(x_s, x_t),
\end{multline*}
where $\newlag^*$ is the vector of Lagrange multipliers with
components 
\[
\newlag^*_{ts}(x_s) \defn \logmess^*_{ts}(x_s) - \sum_{\{ \treegr \,
\big | \, \pa(s) = t\}} \treedist(\treegr) \log \grmaxmarg^*_s(x_s).
\]

\subsection{Proof of Lemma~\ref{LemFinal}}
\label{AppFinal}

Since by assumption the pseudo-max-marginals $\grmaxmarg^*$ defined by
$\mess^* = \exp(\lagmul^*)$ (with the exponential defined
element-wise) satisfy the optimum specification criterion, we can find
a configuration $\mxvec$ that satisfies the local optimality
conditions~\eqref{EqnLocalOpt} for every node and edge on the full
graph $\graph$.  

Since the pseudo-max-marginals $\grmaxmarg^*$ are defined by a fixed
point $\mess^* = \exp(\lagmul^*)$ (with the exponential defined
element-wise) of the update equation~\eqref{EqnMessFix}, the
pseudo-max-marginals must be pairwise-consistent.  More explicitly,
for any edge $(s,t)$ and $x_s \in \statesp_s$, the pairwise
consistency condition $\max_{x_t} \grmaxmarg^*_{st}(x_s, x_t) =
\alphnorm_{st} \; \grmaxmarg^*(x_s)$ holds, where $\alphnorm_{st}$ is
a positive constant independent of $x_s$.  Using this fact, we can
write
\begin{eqnarray}
\label{Eq1}
\max_{x_s, x_t} \log \frac{\grmaxmarg^*_{st}(x_s,
x_t)}{\grmaxmarg^*_s(x_s)} & = & \max_{x_s} \log \frac{ \max_{x_t}
\grmaxmarg^*_{st}(x_s, x_t)}{\grmaxmarg^*_s(x_s)} \nonumber \\ 
& = & \log \alphnorm_{st}.
\end{eqnarray}
Moreover, since by assumption the pseudo-max-marginals $\grmaxmarg^*$
satisfy the optimum specification criterion, we can find a
configuration $\mxvec$ that satisfies the local optimality
conditions~\eqref{EqnLocalOpt} for every node and edge on the full
graph $\graph$.  For this configuration, we have the equality
\begin{eqnarray}
\label{Eqn2a}
\max_{x_s} \log \grmaxmarg^*_s(x_s) = \log \grmaxmarg^*_s(\xs_s)
\qquad \mbox{for all $s \in \vertex$},
\end{eqnarray}
and
\begin{eqnarray}
\label{Eqn2b}
\log \frac{\grmaxmarg_{st}^*(\xs_s, \xs_t)}{\grmaxmarg_s^*(\xs_s)} & =
& \log \frac{ \max_{x_t} \grmaxmarg_{st}^*(\xs_s,
x_t)}{\grmaxmarg_s^*(\xs_s)} \nonumber \\
& = & \log \kappa_{st} \nonumber \\
& = & \max_{x_s, x_t} \log
\frac{\grmaxmarg^*_{st}(x_s, x_t)}{\grmaxmarg^*_s(x_s)}
\end{eqnarray}
for all $(s,t) \in \edge$, where the final equality follows from
equation~\eqref{Eq1}.

Recall the definition of $\taupar^*$ from
equation~\eqref{EqnDefnTaustar}.  Using equations~\eqref{Eqn2a}
and~\eqref{Eqn2b}, we have

\begin{eqnarray}
\label{EqnInqa}
\sum_{x_s} \taupar_s(x_s) \log \grmaxmarg^*_s(x_s) & \leq & \sum_{x_s}
\taupar^*_s(x_s) \log \grmaxmarg^*_s(x_s)
\end{eqnarray}
for all $s \in \vertex$, and
\begin{multline}
\label{EqnInqb}
\sum_{x_s, x_t} \taupar_s(x_s, x_t) \log \frac{\grmaxmarg^*_{st}(x_s,
x_t)}{\grmaxmarg^*_s(x_s)} \\
\leq \sum_{x_s, x_t} \taupar^*_s(x_s, x_t) \log \frac{
\grmaxmarg^*_{st}(x_s,x_t)}{\grmaxmarg^*_s(x_s)}
\end{multline}
for all $(s,t) \in \edge$.  Both equations~\eqref{EqnInqa}
and~\eqref{EqnInqb} hold for all $\taupar \in \Normset$ (i.e., for all
non-negative $\taupar$ such that $\sum_{x_s} \taupar_s(x_s) = 1$ and
$\sum_{x_s, x_t} \taupar_{st}(x_s, x_t) = 1$).  Finally,
inequalities~\eqref{EqnInqa} and~\eqref{EqnInqb} imply that
$\max_{\taupar \in \Normset} \Fcon(\taupar, \grmaxmarg^*; \treegr) =
\Fcon(\taupar^*, \grmaxmarg^*; \treegr)$ as claimed.



{\footnotesize { \bibliographystyle{plain}

}}
\end{document}